\documentclass[aps,pre,twocolumn,groupedaddress,showpacs]{revtex4-2}
\newcommand{\bec}[1]{\mbox{\boldmath $ #1$}}
\usepackage{bm}
\usepackage{graphicx}
\usepackage{color}

{}

{}
{}

\newcommand{\meanN}{\overline{n}}

\newcommand{\meanT}{\overline{T}}

\begin{document}
\title{Experimental study of turbulent thermal diffusion of particles in inhomogeneous and anisotropic turbulence}
\author{E.~Elmakies}
\author{O.~Shildkrot}
\author{N. Kleeorin}
\author{A. Levy}
\author{I.~Rogachevskii}
\email{gary@bgu.ac.il}
\author{A.~Eidelman}

\bigskip
\affiliation{
The Pearlstone Center for Aeronautical Engineering
Studies, Department of Mechanical Engineering,
Ben-Gurion University of the Negev, P.O.Box 653,
Beer-Sheva 8410530,  Israel}

\date{\today}
\begin{abstract}
We study experimentally turbulent thermal diffusion of small particles in inhomogeneous and anisotropic stably stratified turbulence produced by one oscillating grid in the air flow. The velocity fields have been measured using a Particle Image Velocimetry (PIV). We have determined various turbulence characteristics: the mean and turbulent velocities, two-point correlation functions of the velocity field and an integral scale of turbulence from the measured velocity fields. The temperature field have been measured with a temperature probe equipped with 12 E thermocouples. Spatial distributions of micron size particles have been determined by a PIV system using the effect of the Mie light scattering by particles in the flow. The experiments have demonstrated that particles are accumulated at the minimum of mean fluid temperature due to phenomenon of turbulent thermal diffusion.
Using measured spatial distributions of particles and temperature fields, we have determined
the effective turbulent thermal diffusion coefficient
of particles in inhomogeneous temperature stratified turbulence.
This experimental study has clearly detected phenomenon of turbulent thermal diffusion
in inhomogeneous turbulence.
\end{abstract}

\maketitle
\section{Introduction}
\label{sect1}

Turbulent transport and mixing of aerosols and droplets is of
fundamental importance in a large variety of applications ranging from environmental sciences, physics of the
atmosphere and meteorology to industrial turbulent flows and turbulent combustion
\cite{CSA80,ZRS90,BLA97,SP06,ZA08,CST11,RI21}.
Various laboratory experiments and numerical simulations as well as observations in
atmospheric and astrophysical  turbulence have detected large-scale long-living
clusters of particles as well as small-scale particle clusters
\cite{SP06,ZA08,CST11,RI21,WA00,S03,KPE07,G08,WA09}.
Characteristic scales of large-scale clusters are much larger than the integral turbulence scale,
while characteristic scales of small-scale clusters
are much smaller than the integral turbulence scale.

Turbulent diffusion causes a decay of inhomogeneous
particle clusters.
On the other hand, turbulence can create inhomogeneous particle spatial distributions.
For instance, small-scale clusters are formed in non-stratified \cite{BB07,AC08,TB09,BE10,EKR96a,EKR02}
and stratified turbulence \cite{AC08,EKR10,EKR13}.
The large-scale clusters of inertial particles in isothermal non-stratified
inhomogeneous turbulence are caused by turbophoresis
\citep{CTT75,RE83,G97,EKR98,G08,MHR18},
which is a combined effect of particle inertia and inhomogeneity of turbulence.

The large-scale clusters in a temperature-stratified turbulence are formed due to
turbulent thermal diffusion \cite{EKR96,EKR97},
resulting in additional turbulent non-diffusive flux of particles
directed to the minimum of the mean temperature.
The characteristic spatial scale of particle clusters
formed due to turbulent thermal diffusion is much larger than the integral scale of turbulence, and
the characteristic time scale of the formation of the particle clusters is much
larger than the characteristic turbulent time scale.
Turbulent thermal diffusion is a purely collective phenomenon resulting in a pumping effect, described in terms of
effective velocity of particles in the direction opposite to the mean temperature gradient.
A balance between the turbulent thermal diffusion and
turbulent diffusion determines the conditions for the formation of large-scale particle clusters.

Turbulent thermal diffusion has been studied theoretically
\cite{EKR96,EKR97,EKR00,EKRS00,EKRS01,PM02,RE05,AEKR17}
and detected in different laboratory experiments in stably and convective temperature-stratified turbulence
produced by oscillating grids \cite{BEE04,EEKR04,EEKMR06} or a multi-fan generator \cite{EEKR06}.
This phenomenon has been also detected in direct numerical
simulations \cite{HKRB12,RKB18}.
Turbulent thermal diffusion is shown to be a crucial importance
in the atmospheric turbulence with temperature inversions \cite{SSEKR09} and
in astrophysical temperature stratified turbulent flows \cite{H16}.
Turbulent thermal diffusion plays an important role in formation of small-scale particle clusters in turbulence
with a mean vertical temperature gradient \cite{EKR10,EKR13}.

Turbulent thermal diffusion of small particles
has been investigated mainly in a homogeneous turbulence
produced by two oscillating grids \cite{BEE04,EEKR04,EEKMR06,EEKR06}.
In the present experimental study we investigate phenomenon of turbulent
thermal diffusion of small particles in inhomogeneous and anisotropic
stably stratified turbulence produced by one oscillating grid in the air flow.
Previous experiments \cite{turn68,turn73,tho75,hop76,kit97,san98,med01}
with one oscillating grid have been performed in a water flow with isothermal turbulence.
It has been shown in these experiments that
the integral scale $\ell_0$ of turbulence is proportional to
the distance $Y$ from a grid ($\ell_0 \propto Y$), and the root mean square (r.m.s.) velocity
scales as $\sqrt{\langle {\bf u'}^2 \rangle} \propto f \, Y^{-1}$,
where $f$ is the frequency of the grid oscillations. This
implies that Reynolds numbers and the turbulent diffusion
coefficient in the core flow of inhomogeneous turbulence
are weakly dependent on the distance
from the grid.

In this paper we discuss the results of experimental study of turbulent transport
of small particles in inhomogeneous stably stratified turbulence
in the air flow with an imposed temperature gradient.
This paper is organized as follows.
In Section~\ref{sect2} we discuss physics of the phenomenon of turbulent thermal diffusion.
In Section ~\ref{sect3} we describe the experimental setup and measurements techniques.
In Section ~\ref{sect4} we discuss the experimental results
of particle transport in inhomogeneous
and anisotropic stably stratified turbulence produced by
one oscillating grid.
In this section we determine spatial distributions of turbulence parameters
and find spatial distributions of the mean fluid temperature and
the mean number density of particles.
This allows us to determine the characteristics of turbulent thermal diffusion
of particles in inhomogeneous turbulence.
Finally, conclusions are drawn in Section~\ref{sect5}.

\section{Physics of turbulent thermal diffusion}
\label{sect2}

In this section we discuss the physics of phenomenon of turbulent thermal diffusion.
First, we consider dynamics of small non-inertial particles or gaseous admixtures
in a turbulent fluid flow.
Equation for the evolution of the particle number density $n(t,{\bm x})$
in a compressible fluid velocity field ${\bm U}(t,{\bm x})$ reads
\citep{CH43,AP81}
\begin{eqnarray}
{\partial n \over \partial t} + {\bm \nabla} {\bf \cdot}({\bm U}\, n) = D \, \Delta n ,
\label{D1}
\end{eqnarray}
where $D= k_B \,T/(3\pi \rho \, \nu \, d_{\rm p})$ is the coefficient
of the molecular (Brownian) diffusion of particles,   $d_{\rm p}$ is the particle diameter,
$\nu$  is the kinematic viscosity of the fluid, $T$ and $\rho$  are the fluid temperature and
density and $k_B$ is the Boltzmann constant.
We consider fluid flows with a low Mach number, Ma $=|{\bm U}|/c_{\rm s} \ll 1$, which implies that fluid velocity is much less than the sound speed, $c_{\rm s}$.
In this case the continuity equation for the fluid density can be used in an anelastic approximation,
$\bec{\nabla} {\bf \cdot}(\rho \, {\bm U}) = 0 $,
which takes into account an inhomogeneous distribution of the fluid density.

We study a long-term evolution of the particle number density in spatial scales $L_n$ which are much larger than the integral scale of turbulence $\ell_0$, and during the time scales $t_n$ which are much larger than the turbulent time scales $\tau_0$.
We use a mean-field approach, where all quantities are decomposed into the
mean and fluctuating parts, and the fluctuating parts have zero mean values, i.e., we use the Reynolds averaging.
In particular, the particle number density $n= \overline{n} + n'$, where $\overline{n}=\langle n \rangle$ is the mean particle number density, $n'$ are particle number density fluctuations and $\langle n' \rangle=0$. The angular brackets denote ensemble averaging. Averaging Eq.~(\ref{D1}) over an ensemble of turbulent velocity field, we arrive at the mean-field equation for the particle number density:
\begin{eqnarray}
{\partial \overline{n} \over \partial t} + {\bm \nabla} {\bf \cdot} \langle {\bm u} \, n'  \rangle = D \, \Delta \overline{n} ,
\label{D2}
\end{eqnarray}
where $\langle {\bm u} \, n' \rangle$ is the turbulent flux of particles
and ${\bm u}$ are velocity fluctuations.
Here we consider for simplicity the case when the mean velocity vanishes, i.e., $\overline{\bm U}=0$.

Equation~(\ref{D2}) is not yet closed because we do not know how the particle turbulent flux $\langle {\bm u} n' \rangle$ depends on the mean particle number density $\overline{n}$. To determine the particle turbulent flux, we derive an equation for particle number density fluctuations $n'$, that is obtained by subtracting Eq.~(\ref{D2}) from Eq.~(\ref{D1}):
\begin{eqnarray}
{\partial n' \over \partial t} + {\bm \nabla} {\bf \cdot} \left(n'  {\bm u} - \langle n'  {\bm u} \rangle \right) - D  \Delta n' = -({\bm u} {\bf \cdot} {\bm \nabla}) \overline{n} -\overline{n} ({\bm \nabla} {\bf \cdot} {\bm u}) .
\nonumber\\
\label{D3}
\end{eqnarray}
The term, ${\cal Q} = {\bm \nabla}{\bf \cdot} \left(n' \, {\bm u} - \langle n' \, {\bm u} \rangle \right)$, in the left-hand side of Eq.~(\ref{D3}) is the nonlinear term, while
the terms, $-({\bm u} {\bf \cdot} {\bm \nabla}) \overline{n} -\overline{n} ({\bm \nabla} {\bf \cdot} {\bm u})$, in the right-hand side of Eq.~(\ref{D3}) are the source terms for particle number density fluctuations.
The first source term, $-({\bm u} {\bf \cdot} \bec{\nabla}) \overline{n}$, in Eq.~(\ref{D3})
causes production of particle number density fluctuations by the tangling of the gradient of the mean particle number density ${\bm \nabla} \, \overline{n}$ by velocity fluctuations.

The second source term in Eq.~(\ref{D3}) is $-\overline{n} ({\bm \nabla} {\bf \cdot} {\bm u})$.
Let us decompose the fluid density into the mean fluid density $\overline{\rho}$ and fluctuations $\rho'$, i.e., $\rho= \overline{\rho}  + \rho'$, where for low Mach numbers $|\rho'| \ll \overline{\rho}$.
The anelastic approximation yields $\bec{\nabla} {\bf \cdot} (\rho \, {\bm u}) \approx \bec{\nabla} {\bf \cdot} (\overline{\rho} \, {\bm u}) = 0$, so that
$\bec{\nabla} {\bf \cdot} {\bm u} \approx - ({\bm u} \cdot {\bm \nabla}) \,\overline{\rho} / \overline{\rho}$.
Introducing a vector ${\bm \lambda} = - {\bm \nabla} \overline{\rho} / \overline{\rho}$,
we obtain that $\bec{\nabla} {\bf \cdot} {\bm u} \approx  {\bm u} \cdot {\bm \lambda}$.
Thus, the second source term, $-\overline{n} ({\bm \nabla} {\bf \cdot} {\bm u}) = (\overline{n}/\overline{\rho} \,) ({\bm u} {\bf \cdot} {\bm \nabla}) \overline{\rho} = - \overline{n} \, ({\bm u} {\bf \cdot} {\bm \lambda})$, causes production of particle number density fluctuations by the tangling of the gradient of the mean fluid density ${\bm \nabla} \, \overline{\rho} \, $ by velocity fluctuations.
The ratio of the absolute values of the nonlinear term $|{\cal Q}|$ to the diffusion term $|D \Delta n'|$ is the P\'{e}clet number for particles,
that can be estimated as ${\rm Pe} = u_0 \, \ell_0 / D$.

Equation~(\ref{D3}) is a nonlinear equation for particle number density fluctuations. Since this nonlinear equation cannot be solved exactly for arbitrary P\'{e}clet numbers, one has to use different approximate methods for the solution of Eq.~(\ref{D3}). We consider a one-way coupling, i.e., we take into account the effect of the turbulent velocity on the particle number density, but we neglect the feedback effect of the particle number density on the turbulent fluid flow. This approximation is valid when the spatial density of particles $n \, m_p$ is much smaller than the fluid density $\rho$, where $m_p$ is the particle mass.
We also consider non-inertial particles or gaseous admixtures. In this case the particles move with the fluid velocity. These assumptions imply that the particle number density is a passive scalar.

We use for simplicity the dimensional analysis to solve Eq.~(\ref{D3}). The dimension of
the left-hand side of Eq.~(\ref{D3}) is the rate of change of particle number density fluctuations $n' / \tau_{n'}$, where $\tau_{n'}$ is the characteristic time of particle number density fluctuations.
For large Reynolds and P\'{e}clet numbers, the characteristic time of particle number density fluctuations $\tau_{n'}$
can be identified with the correlation time $\tau_0$ of the turbulent velocity field.
Therefore, in the framework of the dimensional analysis, we replace the left-hand side of Eq.~(\ref{D3}) by $n' / \tau_0$. This yields:
\begin{eqnarray}
n' = - \tau_0 \, \left[({\bm u} {\bf \cdot} {\bm \nabla}) \overline{n} + \overline{n} \, ({\bm \nabla} {\bf \cdot} {\bm u})\right].
\label{D4}
\end{eqnarray}
Multiplying Eq.~(\ref{D4}) by velocity fluctuations, $u_i$, and averaging
over an ensemble of turbulent velocity field, we arrive at the expression for the turbulent flux of particles:
\begin{eqnarray}
\left\langle n' \, u_i \right\rangle &=& - \tau_0 \,\left\langle u_i u_j  \right\rangle \, \nabla_j \overline{n} - \tau_0 \, \overline{n} \,\left\langle u_i (\bec{\nabla} {\bf \cdot} {\bm u}) \right\rangle
\nonumber\\
&\equiv& V_i^{\rm eff} \, \overline{n} - D_{ij}^{(n)} \, \nabla_j \overline{n} ,
\label{D5}
\end{eqnarray}
where the last term, $- D_{ij}^{(n)} \, \nabla_j \overline{n} $, in
Eq.~(\ref{D5}) determines the contribution to the flux of particles caused by
turbulent diffusion, and
$D_{ij}^{(n)} = \tau_0 \,\left\langle u_i u_j  \right\rangle$
is the turbulent diffusion tensor.
For an isotropic turbulence
$\langle u_i u_j  \rangle = \delta_{ij} \, \langle {\bm u}^2  \rangle/3$,
so that the turbulent diffusion tensor for large P\'{e}clet numbers is given by
$D_{ij}^{(n)} =D_{\rm T} \delta_{ij}$, where $D_{\rm T}= \tau_0 \,\left\langle {\bm u}^2  \right\rangle / 3$
is the turbulent diffusion coefficient.

The term ${\bm V}^{\rm eff} \, \overline{n}$ in
Eq.~(\ref{D5}) determines the contribution to the turbulent flux of particles caused by
the effective pumping velocity:
${\bm V}^{\rm eff} = - \tau_0 \,\left\langle {\bm u} (\bec{\nabla} {\bf \cdot} {\bm u}) \right\rangle$.
Next, we take into account the anelastic approximation,
${\bm \nabla} {\bf \cdot} {\bm u} = - (1/\overline{\rho}) \, ({\bm u} {\bf \cdot} {\bm \nabla}) \overline{\rho}$, so that
the effective pumping velocity is given by
$V_i^{\rm eff} = - \tau_0 \, \left\langle u_i u_j \right \rangle \, \lambda_j  $, where
${\bm \lambda} = - {\bm \nabla} \overline{\rho} / \overline{\rho}$.
For an isotropic turbulence, the effective pumping velocity is given by \cite{EKR96,EKR97}
\begin{eqnarray}
{\bm V}^{\rm eff}  = D_{\rm T} \, {{\bm \nabla} \overline{\rho} \over \overline{\rho}},
\label{D7}
\end{eqnarray}
and the particle turbulent flux $\langle  {\bm u} \, n'  \rangle$ is
\begin{eqnarray}
\left\langle {\bm u} \, n'  \right\rangle = {\bm V}^{\rm eff} \, \overline{n} - D_{\rm T} \, {\bm \nabla} \overline{n} .
\label{D8}
\end{eqnarray}
To understand the physics related to the effective pumping velocity ${\bm V}^{\rm eff}$, let us first
express the effective pumping velocity via physical parameters.
We use the equation of state for a perfect gas,
$P=(k_B/ m_\mu) \, \rho \, T$, that can be also rewritten for the mean fields as
$\overline{P}=(k_B/ m_\mu) \, \overline{\rho} \, \overline{T}$,
where $\overline{P}$ and $\meanT$ are the mean pressure
and mean temperature, respectively.
Here we assume that $\overline{\rho} \, \meanT \gg \langle \rho' \, \theta \rangle$.
By means of the equation of state,
we express the gradient of the mean fluid density in terms of the gradients of
the mean fluid pressure ${\bm \nabla} \overline{P}$ and mean fluid temperature ${\bm \nabla} \meanT$ as
${\bm \nabla} \, \ln\overline{\rho} = {\bm \nabla} \ln \overline{P} - {\bm \nabla} \ln \meanT$.
For small mean pressure gradient, ${\bm \nabla} \, \ln\overline{\rho} \approx  - {\bm \nabla} \meanT$, so that
the effective pumping velocity of non-inertial particles is given by \cite{EKR96,EKR97}
\begin{eqnarray}
{\bm V}^{\rm eff} = - D_{\rm T} \, {{\bm \nabla} \meanT \over \meanT} .
\label{D9}
\end{eqnarray}
The rigorous methods yield the result similar to Eq.~(\ref{D9}) (see, e.g., \cite{RI21}).
For inertial particles, the effective pumping velocity is given by
\begin{eqnarray}
{\bm V}^{\rm eff} = - \alpha\left(d_{\rm p}, {\rm Re} \right) \, D_{\rm T} \, {{\bm \nabla} \meanT \over \meanT} ,
\label{D10}
\end{eqnarray}
where the effective turbulent thermal diffusion coefficient of particles is
\begin{eqnarray}
 \alpha(d_{\rm p}, {\rm Re}) =1 + {\tau_p(d_{\rm p})  \over \tau_0} \, {\rm Re}^{1/4} \, \ln({\rm Re})  \, \left({L_{\rm eff} \over \ell_0}\right) ,
 \nonumber\\
\label{BBB9}
\end{eqnarray}
(see \cite{EKR98,EKR00,EKR13,AEKR17}). Here
$L_{\rm eff} = 2 c_s^2 \tau_\eta^{3/2}/3 \nu^{1/2}$ is the effective length scale, $c_s$ is the
sound speed,  $\tau_{\rm p} = m_{\rm p} / (3 \pi \rho \, \nu d_{\rm p})$ is the Stokes time
for particles having mass $m_{\rm p}$ and diameter $d_{\rm p}$, $\tau_\eta=\tau_0 / \sqrt{\rm Re}$
is the Kolmogorov viscous time, ${\rm Re} = \tau_0 \langle {\bm u}^2  \rangle / \nu$
is the Reynolds number and $\nu$ is the kinematic viscosity.
For the conditions pertinent to our laboratory experiments, the function $\alpha$ varies
for the micron-size particles from 2 to 3 depending on Reynolds number.

Substituting Eq.~(\ref{D8}) into Eq.~(\ref{D2}), we arrive at the evolutionary equation for the particle mean number density as
\begin{eqnarray}
{\partial \meanN \over \partial t} + \nabla_z \left[V_z^{\rm (eff)} \, \meanN - \left(D+D_{\rm T}\right) \,\nabla_z \meanN \right]=0 .
  \label{D11}
\end{eqnarray}
The steady-state solution of Eq.~(\ref{D11}) for the  mean number density of particles with
a zero total flux of particles at the vertical boundaries is given by
\begin{eqnarray}
{\meanN(z)  \over  \meanN_0} = \left({\meanT(z)  \over \meanT_0}\right)^{- {\alpha D_{\rm T} \over D_{\rm T}+D}} ,
 \label{D12}
\end{eqnarray}
where $\meanT_0$ and $\meanN_0$ are the values of the mean fluid temperature and the particle mean number density
at the vertical boundaries.
Here we take into account Eq.~(\ref{D10}). It follows from Eq.~(\ref{D12}), that particles are accumulated
at the vicinity of the minimum of the mean temperature.

The physics of the effect of turbulent thermal diffusion
for solid particles is as follows \cite{EKR96,EKR97}.
The inertia causes particles inside the turbulent eddies to drift
out to the boundary regions between eddies due to the centrifugal inertial force,
so that inertial particles are locally accumulated in these regions.
These regions have low vorticity fluctuations, high strain rate and high pressure fluctuations \cite{M87}.
Similarly, there is an outflow of inertial particles from regions with minimum fluid
pressure fluctuations.
In homogeneous and isotropic turbulence with a zero gradient
of the mean temperature, there is no preferential direction,
so that there is no large-scale effect of particle accumulation.

In temperature-stratified turbulence, fluctuations of fluid temperature $\theta$
and velocity ${\bm u}$ are correlated due to a
non-zero turbulent heat flux, $\langle
\theta \, {\bm u} \rangle\not=\bm{0}$. Fluctuations of
temperature cause pressure fluctuations, which
result in fluctuations of the number density of
particles.
In the mechanism of turbulent thermal diffusion, only pressure fluctuations which
are correlated with velocity fluctuations
due to a non-zero turbulent heat flux play a crucial role.
Increase of the fluid pressure fluctuations
is accompanied by an accumulation of particles,
and the direction of the mean flux of particles
coincides with that of the turbulent heat flux.
The turbulent flux of particles is directed to the
minimum of the mean temperature, and the
particles tend to be accumulated in this
region \cite{RI21}.

The similar effect of accumulation of particles
in the vicinity of the mean temperature minimum
(or in the vicinity of the
maximum of the mean fluid density) and the formation of
inhomogeneous spatial distributions of the mean particle number density
exists also for non-inertial particles or gaseous admixtures
in density-stratified or temperature-stratified turbulence, i.e.,
for a low-Mach-number compressible turbulent fluid flow \cite{EKR97,HKRB12,RKB18}.

The physics of the accumulation
of non-inertial particles in the vicinity of the
maximum of the mean fluid density can be explained
as follows.
Let us assume that the mean fluid density
$\overline{\rho}_2$ at point $2$ is larger than the mean
fluid density $\overline{\rho}_1$ at point $1$. Consider
two small control volumes {\it a} and
{\it b} located between these two points,
and let the direction of the local turbulent
velocity in volume {\it a} at some instant
be the same as the direction of the mean fluid
density gradient $\bec\nabla \, \overline{\rho}$ (i.e.,
along the $x$ axis toward point $2$). Let the
local turbulent velocity in volume {\it b}
at this instant be directed opposite to the mean
fluid density gradient (i.e., toward point $1$).

In a fluid flow with a nonzero mean fluid density
gradient, one of the sources of particle number
density fluctuations, $n' \propto - \tau_0 \, \overline{n} \,
(\bec\nabla {\bf \cdot} \, {\bm u})$, is caused by a
non-zero $\bec\nabla\cdot{\bm u} \approx - {\bm u} \cdot
\bec\nabla \ln \overline{\rho} \not=0$.
Since fluctuations of the fluid velocity ${\bm u}$ are
positive in volume {\it a} and negative in
volume {\it b}, we have the negative divergence of the fluid velocity, $\bec\nabla {\bf
\cdot} \, {\bm u} < 0$, in volume {\it a}, and the positive divergence of the fluid velocity,
$\bec\nabla {\bf \cdot} \, {\bm u} > 0$, in volume
{\it b}.
Therefore, fluctuations of the
particle number density $n' \propto - \tau_0 \, \overline{n}
\, (\bec\nabla {\bf \cdot} \, {\bm u})$ are positive
in volume {\it a} and negative in volume
{\it b}. However, the flux of particles $n'\,
u_x$ is positive in volume {\it a} (i.e., it
is directed toward point $2$), and it is also
positive in volume {\it b} (because both
fluctuations of fluid velocity and number density
of particles are negative in volume
{\it b}). Therefore, the mean flux of
particles $\langle n' {\bm u} \rangle$ is directed, as is
the mean fluid density gradient $\bec\nabla \,
\overline{\rho}$, toward point~2. This results in formation large-scale
heterogeneous structures of non-inertial
particles in regions with a mean fluid density
maximum. When the gradient $\bec\nabla \,
\overline{P}$ of the mean fluid pressure  vanishes, $(\bec\nabla \, \overline{\rho}) /\overline{\rho}
= - (\bec\nabla \, \overline{T})/\overline{T}$.
This implies that particles are accumulated at the vicinity
of the minimum of the fluid temperature  \citep{EKR97,RI21}.

Compressibility in a low-Mach-number stratified turbulent fluid flow
causes an additional non-diffusive component of the turbulent flux of non-inertial
particles or gases, and results in the formation of large-scale inhomogeneous structures
in spatial distributions of non-inertial particles.
In a temperature stratified turbulence, preferential concentration of particles
caused by turbulent thermal diffusion
can occur in the vicinity of the minimum of the mean temperature.

\section{Experimental setup}
\label{sect3}

\begin{figure}
\centering
\includegraphics[width=7.5cm]{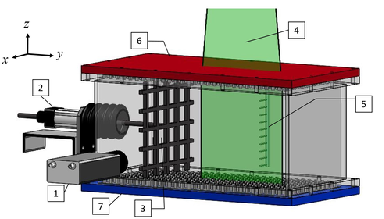}
\caption{\label{Fig1}
Experimental setup: (1) digital CCD camera; (2) rod driven by the
speed-controlled motor; (3) oscillating grid;  (4) laser light sheet;
(5) temperature probe equipped with 12 E - thermocouples;
(6) heat exchanger at the top cooled wall of the chamber;
(7) heat exchanger at the bottom heated wall of the chamber.
}
\end{figure}

In the present paper we study turbulent
thermal diffusion of small particles in experiments with inhomogeneous and anisotropic
stably stratified turbulence produced by one oscillating grid in the air flow.
In this section we describe very briefly the experimental set-up
and measurement facilities in the oscillating grid turbulence.
The experiments in stably stratified
turbulence have been conducted in rectangular
chamber.
The dimensions of the chamber are $L_x \times L_y \times L_z$, where
$L_x=L_z=26$ cm, $L_y=53$ cm.  The axis $Z$ is in the vertical
direction and the axis $Y$ is perpendicular to the grid plain, so that the distance from the grid
is measured along the $Y$ axis.
Turbulence in a transparent chamber is produced by one oscillating
vertically oriented grid with bars arranged in a square array.
The grid is parallel to the side
walls and positioned at a distance of two grid meshes from
the chamber left wall (see Fig.~~\ref{Fig1}). The grid is operated at the frequency $10.5$ Hz.
The chosen frequency applied to the grid is the maximum possible frequency
in our experimental setup before the grid can be damaged.
It allows us to produce turbulence with the maximum intensity.

A vertical mean temperature gradient in the
turbulent air flow is formed by attaching two
aluminium heat exchangers to the bottom (cooled) and top (heated)
walls of the chamber which allow us to
form a mean temperature gradient in a turbulent flow.
To improve the heat transfer in the boundary layers at the bottom and top walls
of the chamber, we use the heat exchangers with rectangular pins $3 \times 3 \times 15$
mm. This allows us to support a large mean temperature gradient in the
core of the flow.

The temperature field is measured with  a temperature probe equipped with 12
E - thermocouples (with the diameter of 0.13 mm and
the sensitivity of $\approx 75 \, \mu$V/K)
attached to a vertical rod with a diameter 4 mm.
The mean spacing between thermal couples along the rod
is about 21.6 mm. Each thermocouple is inserted into a
1.15 mm diameter and 36 mm long case. A tip of a
thermocouple protruded at the length of 8 mm out
of the case.

The temperature is measured for 12
rod positions with 20 mm intervals in the
horizontal direction.
A sequence of 530 temperature readings for every
thermocouple at every rod position is recorded
and processed using the developed software based
on LabView 7.0. Temperature maps are obtained
using Matlab~9.7.0.
A large mean temperature gradient up to 1.2~K/cm in the core flow and 3~K/cm close to the walls
at a mean temperature of about 310~K can be formed.

The velocity field is measured using a
Particle Image  Velocimetry (PIV) \cite{AD91,RWK07,W00}. In the
experiments we use LaVision Flow Master III
system. A double-pulsed light sheet is provided
by a Nd-YAG laser (Continuum Surelite $2 \times
170$ mJ). The light sheet optics includes
spherical and cylindrical Galilei telescopes with
tuneable divergence and adjustable focus length.
We use a progressive-scan 12 bit digital CCD
camera (with pixel size $6.45 \, \mu$m $\times \,
6.45 \, \mu$m and $1376 \times 1040$ pixels) with
a dual-frame-technique for cross-correlation
processing of captured images. A programmable
Timing Unit (PC interface card) generate
sequences of pulses to control the laser, camera
and data acquisition rate. The software package
LaVision DaVis 8.3 is applied to control all
hardware components and for 32 bit image
acquisition and visualization. This software
package comprises PIV software for calculating
the flow fields using cross-correlation analysis.

An incense smoke with sub-micron particles
($\rho_{\rm p} / \rho \sim 10^3)$,  is used as a
tracer for the PIV measurements,
where $\rho_{\rm p}$ is the material density of particles. Smoke is
produced by high temperature sublimation of solid
incense grains. Analysis of smoke particles using
a microscope (Nikon, Epiphot with an
amplification of 560) and a PM-300 portable laser
particulate analyzer shows that these particles
have an approximately spherical shape and that
their mean diameter is of the order of $0.7
\mu$m. The probability density function of the
particle size measured with the PM-300 particulate
analyzer is independent of the location in the
flow for incense particle size of $0.5-1 \,\mu
$m. The maximum tracer particle displacement in
the experiment is of the order of $1/4$ of the
interrogation window. The average displacement of
tracer particles is of the order of $1.5$
pixels. The average accuracy of the velocity
measurements is of the order of $4 \%$ for the
accuracy of the correlation peak detection in the
interrogation window of the order of $0.1$ pixel
\cite{AD91,RWK07,W00}.

We have determined the following turbulence characteristics
in the experiments: the mean and the root mean square (r.m.s.)
velocities, two-point correlation
functions and an integral scale of turbulence
from the measured velocity fields. Series of 530
pairs of images acquired with a frequency of 4
Hz, are stored for calculating velocity maps and
for ensemble and spatial averaging of turbulence
characteristics.  We measure velocity in a flow
domain $197.12 \times 157.7$ mm$^2$ with a spatial
resolution of $1280 \times 1024$ pixels. This
corresponds to a spatial resolution 154 $\mu$m /
pixel. The velocity field in the probed region
is analyzed with interrogation windows of $32
\times 32$ pixels. In every
interrogation window a velocity vector is
determined from which velocity maps comprising
$80 \times 64$ vectors are
constructed. The mean and r.m.s. velocities for
every point of a velocity map are calculated by
averaging over 530 independent maps.

The two-point correlation functions  of the
velocity field are determined for every point of
the velocity map (with $80 \times 64$ vectors) by averaging over 530
independent velocity maps.
The integral scales of turbulence $\ell_y$ and $\ell_z$ are determined
in the horizontal $Y$ and the vertical $Z$ directions from the
two-point correlation functions of the velocity
field.

Particle spatial distribution is determined using PIV system. In particular, the effect of the Mie
light scattering by particles was used to determine the particle
spatial distribution in the flow \cite{guib01}.
The mean intensity of scattered light was
determined in $40 \times 32$ interrogation windows with the size
$32 \times 32$ pixels. The vertical distribution of the intensity of
the scattered light was determined in 40 vertical strips composed of
32 interrogation windows.

The light radiation energy flux scattered
by small particles is $E_s \propto E_0 \Psi(\pi d_{\rm p}/\lambda; a_0;
n)$, where $E_0 \propto \pi d_{\rm p}^2 / 4$ is the energy flux incident
at the particle, $d_{\rm p}$ is the particle diameter, $\lambda$ is the
wavelength, $a_0$ is the index of refraction and $\Psi$ is the
scattering function. For wavelengths $\lambda$ which are larger than
the particle  perimeter $(\lambda > \pi d_{\rm p})$, the function $\Psi$
is given by Rayleigh's law, $\Psi \propto d_{\rm p}^4$. If the wavelength
is small, the function $\Psi$ tends to be independent of $d_{\rm p}$ and
$\lambda$. In the general case the function $\Psi$ is given by the Mie
equations \cite{BH83}.

The scattered light energy flux incident on the CCD camera probe
(producing proportional charge in every CCD pixel) is proportional
to the particle number density $n$, i.e., $ E_s \propto E_0 \, n \,
(\pi d_{\rm p}^2 / 4) $. The probability density function of the particle
size (measured with the PM300 particulate analyzer) was independent
of the location in the flow. Indeed, since the number density of
particles is small, so that they are about $1$ mm apart, it can be
safely assumed that a change in particle number density $n$ does not
affect their size distribution. Consequently, the ratio of the
scattered radiation fluxes at two locations in the flow and at the
image measured with the charge-coupled device (CCD) camera is equal to the ratio of the
particle number densities at these two locations. Measurements
performed using different concentration of the incense smoke showed
that the distribution of the average scattered light intensity over
a vertical coordinate was independent of the particle number
density in the isothermal flow.

\begin{figure}
\centering
\includegraphics[width=8.0cm]{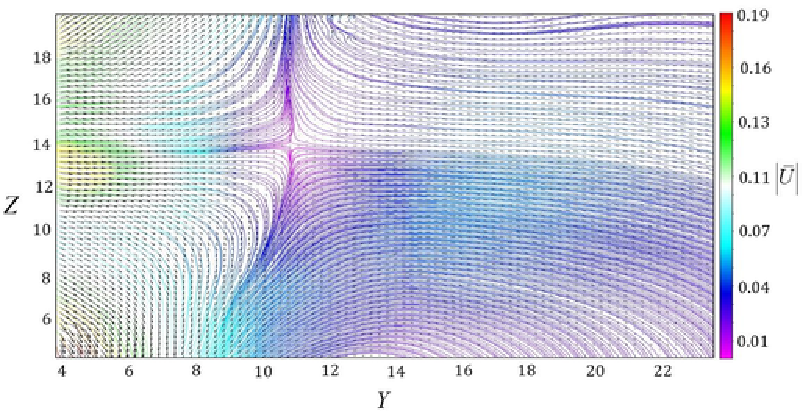}
\caption{\label{Fig2}
Mean velocity field  in the core flow for isothermal turbulence. The velocity is measured
in m/s and coordinates are in cm.
}
\end{figure}

\begin{figure}
\centering
\includegraphics[width=8.0cm]{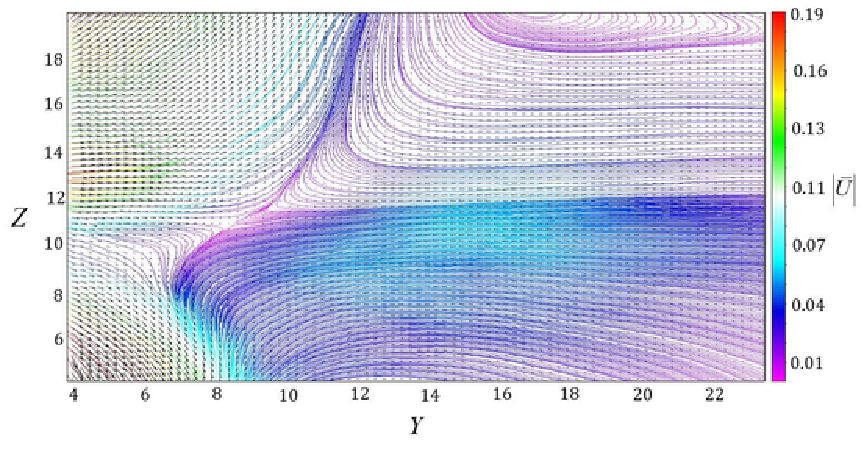}
\caption{\label{Fig3}
Mean velocity field in the core flow for temperature stratified turbulence.
The velocity is measured in m/s and coordinates are in cm.
}
\end{figure}

To characterize the spatial distribution of particle
number density $n \propto E^T /E$ in the non-isothermal flow,
the distribution of the scattered light intensity E measured in the
isothermal case was used for the normalization of the scattered light intensity $E^T$ obtained in
a non-isothermal flow under the same conditions. The
scattered light intensities $E^T$ and $E$ in each experiment
were normalized by corresponding scattered light intensities
averaged over the vertical coordinate. The Mie scattering
is not affected by temperature change because it depends
on the electric permittivity of particles, the particle size
and the laser light wave length. The temperature effect
on these characteristics is negligibly small.

Similar experimental set-up
and data processing procedure have been previously used by us
in the experimental study of different
aspects of turbulent convection \cite{EEKR06,BEKR09,EEKR11},
stably stratified turbulence \cite{EEKR13}, phenomenon of turbulent thermal
diffusion in homogeneous turbulence \cite{BEE04,EEKR04,EEKMR06,EEKR06,AEKR17}
and small-scale particle clustering \cite{EKR10}.

\section{Experimental results}
\label{sect4}

In this section we describe the obtained experimental results.
The experiments for the temperature difference $\Delta T =50$~K
between the top and bottom walls have been performed in the present study.
From the measured velocity fields, we determine various turbulence characteristics for isothermal and stably stratified turbulence
in horizontal and vertical directions: the mean velocity patterns,  the turbulent velocity distributions, the two-point correlation functions of the velocity field which allow us to find the integral scales of turbulence.

\begin{figure}
\centering
\includegraphics[width=7.0cm]{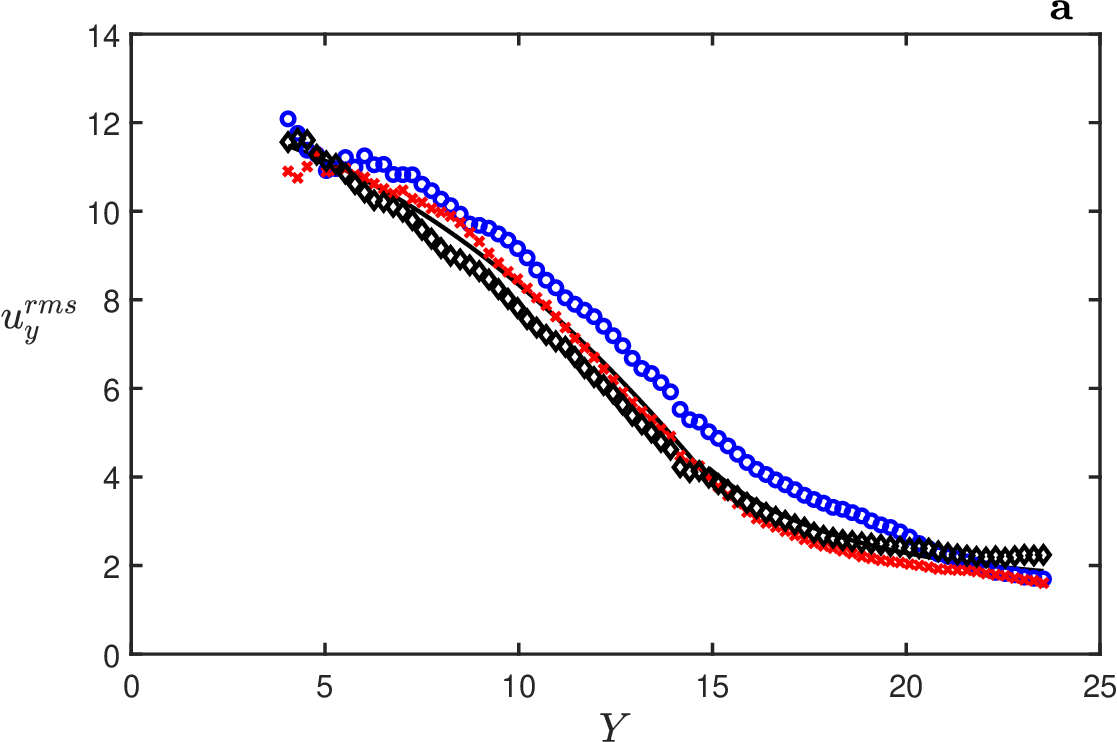}
\includegraphics[width=7.0cm]{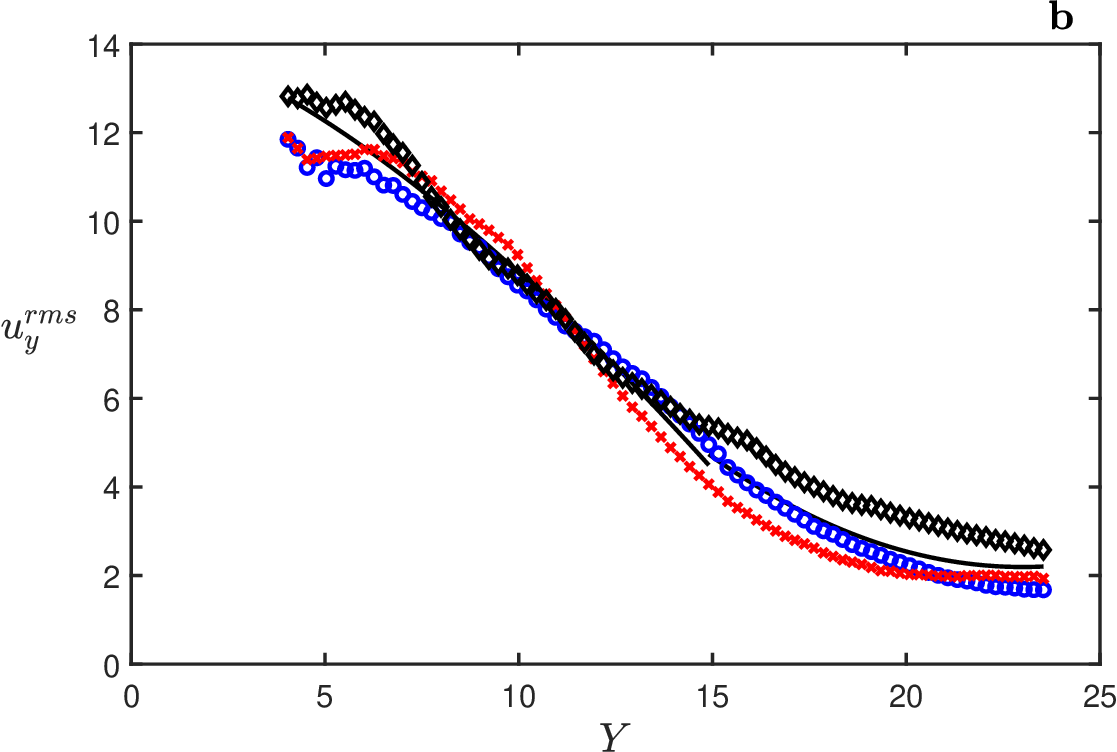}
\caption{\label{Fig4}
Horizontal component of the turbulent velocity $u^{\rm (rms)}_y$ versus $Y$
averaged over different vertical regions: $Z=6.5-11$ cm (black, diamonds); $Z=11-15$ cm (red, slanting crosses); $Z=15.2-18$ cm
(blue, circles) for  {\bf (a)} isothermal turbulence (upper panel); fitting: $u^{\rm (rms)}_y=-0.05Y^{1.8}+12$  (solid)  from $Y =4$ cm to 15 cm and  $u^{\rm (rms)}_y=10^5Y^{-4}+1$ (solid) from $Y =15$ cm to 23.7 cm;  and {\bf (b)} temperature stratified turbulence
(lower panel); fitting: $u^{\rm (rms)}_y=-0.17Y^{1.496}+14.15$ (solid) from $Y =4$ cm to 15 cm, $u^{\rm (rms)}_y=0.038 \, Y^{-1.786}+22.73$ (solid) from $Y =15$ cm to 23.7 cm.
The velocity is measured in cm/s and coordinates  are in cm.
}
\end{figure}

\begin{figure}
\centering
\includegraphics[width=7.0cm]{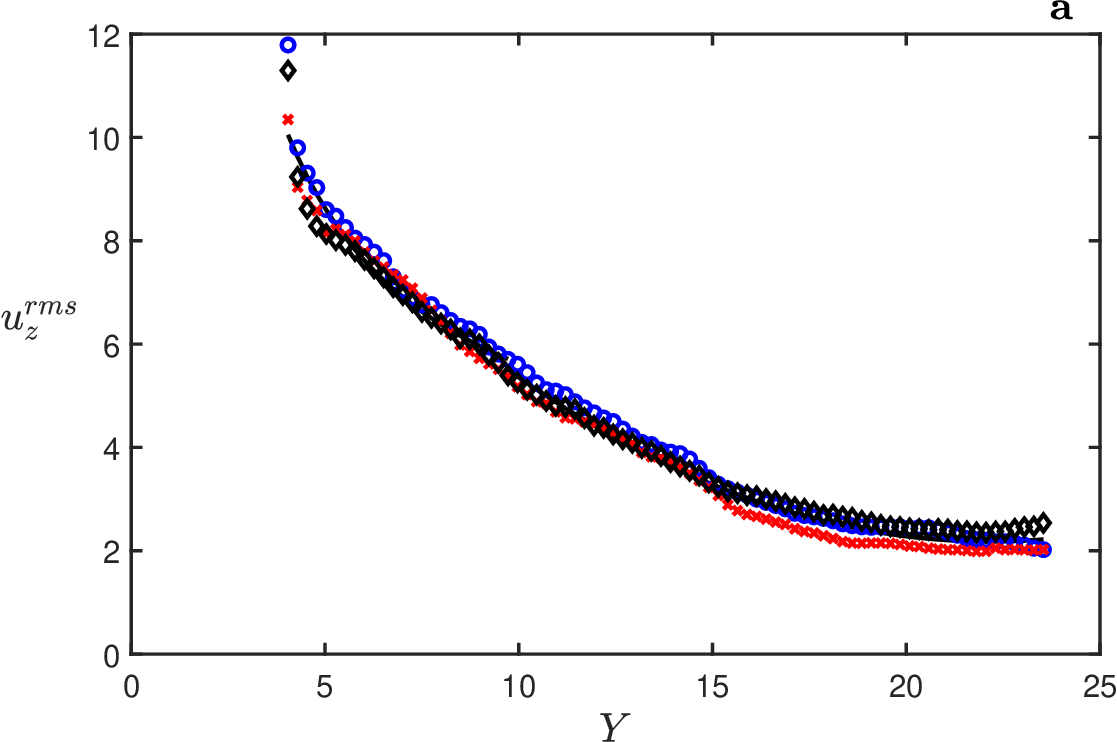}
\includegraphics[width=7.0cm]{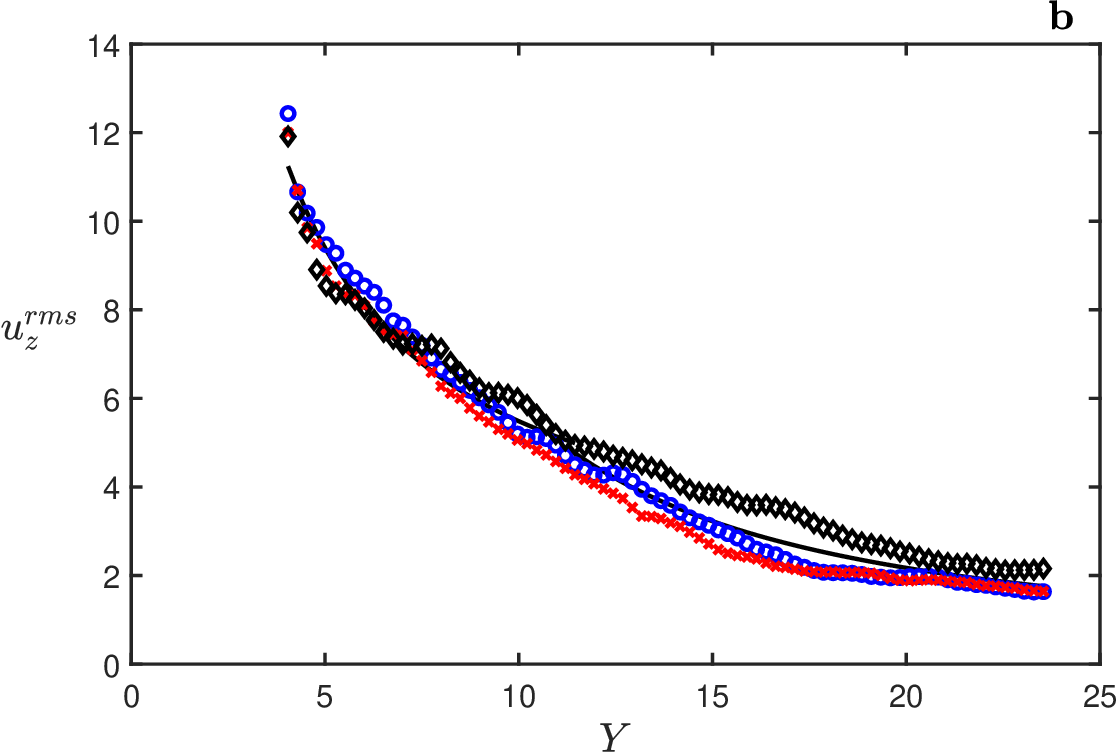}
\caption{\label{Fig5}
Vertical component of the turbulent velocity $u^{\rm (rms)}_z$ versus $Y$
averaged over different vertical regions: $Z=6.5-11$ cm (black, diamonds); $Z=11-15$ cm (red, slanting crosses); $Z=15.2-18$ cm
(blue, circles) for {\bf (a)} isothermal turbulence (upper panel);
fitting:  $u^{\rm (rms)}_z=33.58Y^{-1}+2$ (solid) from $Y=4$ cm to 10 cm, and $u^{\rm (rms)}_z=0.02Y^{2}-0.95Y+12.7$ (solid) from $Y=10$ cm to 23.7 cm;
and {\bf (b)} temperature stratified turbulence
(lower panel); fitting: $u^{\rm (rms)}_z=42.8Y^{-1}+1$ (solid) from $Y =4$ cm to 11.5 cm; and $u^{\rm (rms)}_z=213Y^{-1.6}+0.36$ (solid)  from  $Y =11.5$ cm to 23.7 cm.
The velocity is measured in cm/s and coordinates are in cm.
}
\end{figure}

\begin{figure}
\centering
\includegraphics[width=7.0cm]{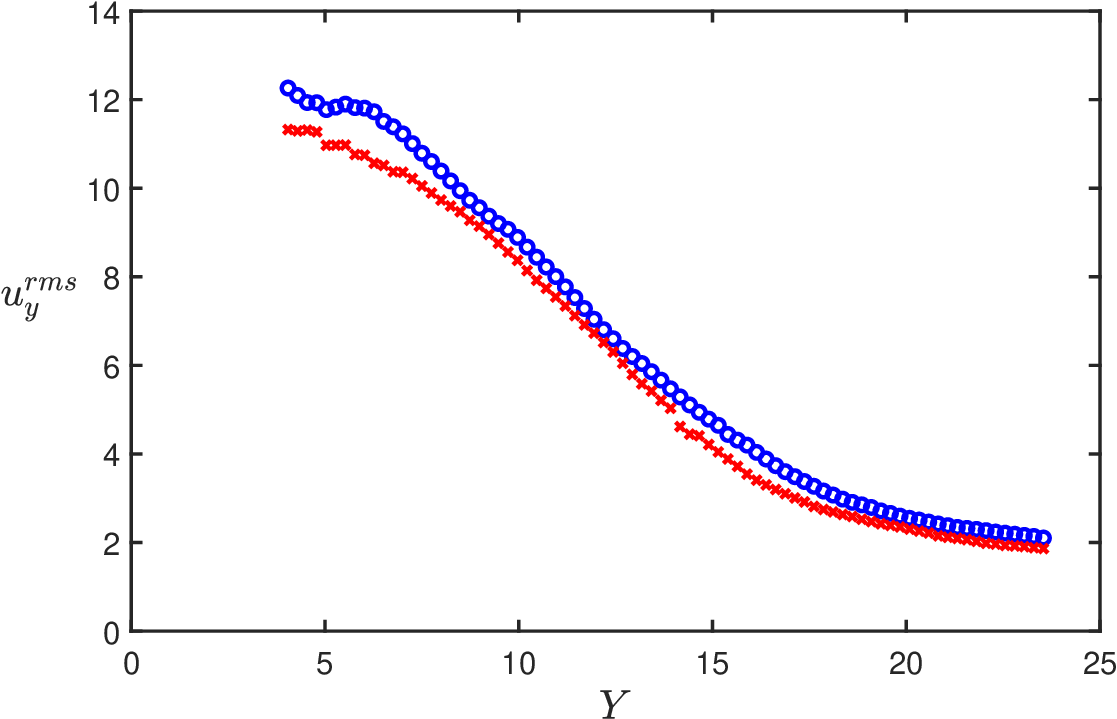}
\caption{\label{Fig6}
Horizontal component of the turbulent velocity $u^{\rm (rms)}_y$ versus $Y$
averaged over $Z$ for isothermal turbulence (red) and temperature stratified turbulence (blue).
The velocity is measured in cm/s and coordinates  are in cm.
}
\end{figure}

\begin{figure}
\centering
\includegraphics[width=7.0cm]{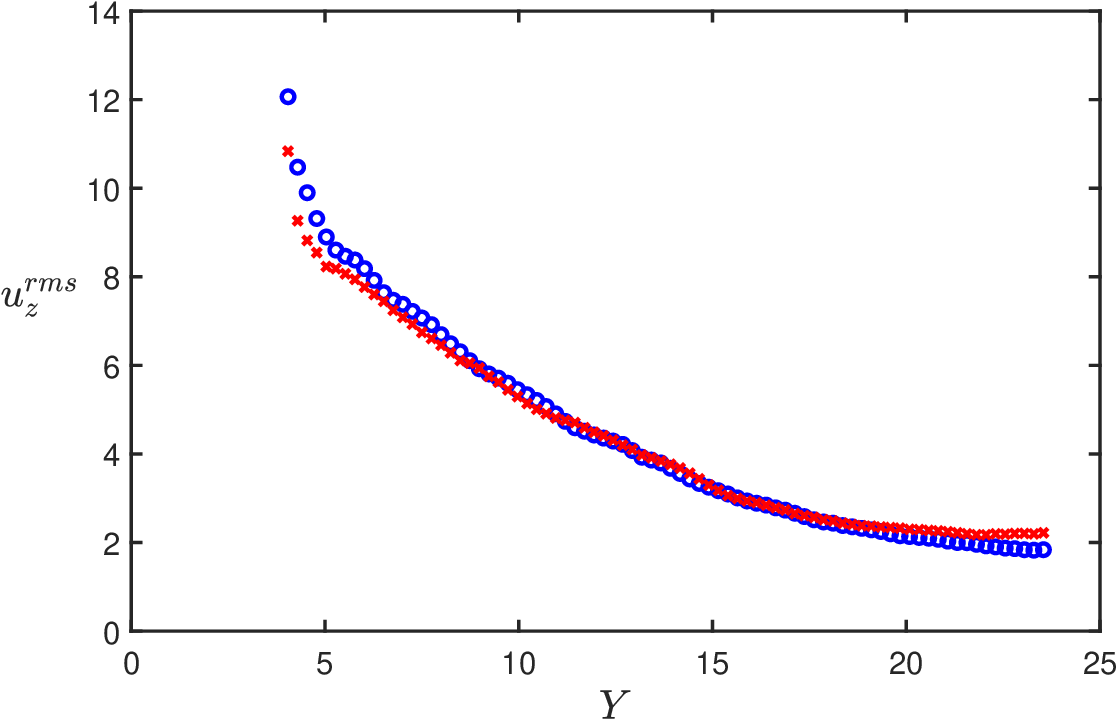}
\caption{\label{Fig7}
Horizontal component of the turbulent velocity $u^{\rm (rms)}_z$ versus $Y$
averaged over $Z$ for isothermal turbulence (red) and temperature stratified turbulence (blue).
The velocity is measured in cm/s and coordinates  are in cm.
}
\end{figure}

\begin{figure}
\centering
\includegraphics[width=7.0cm]{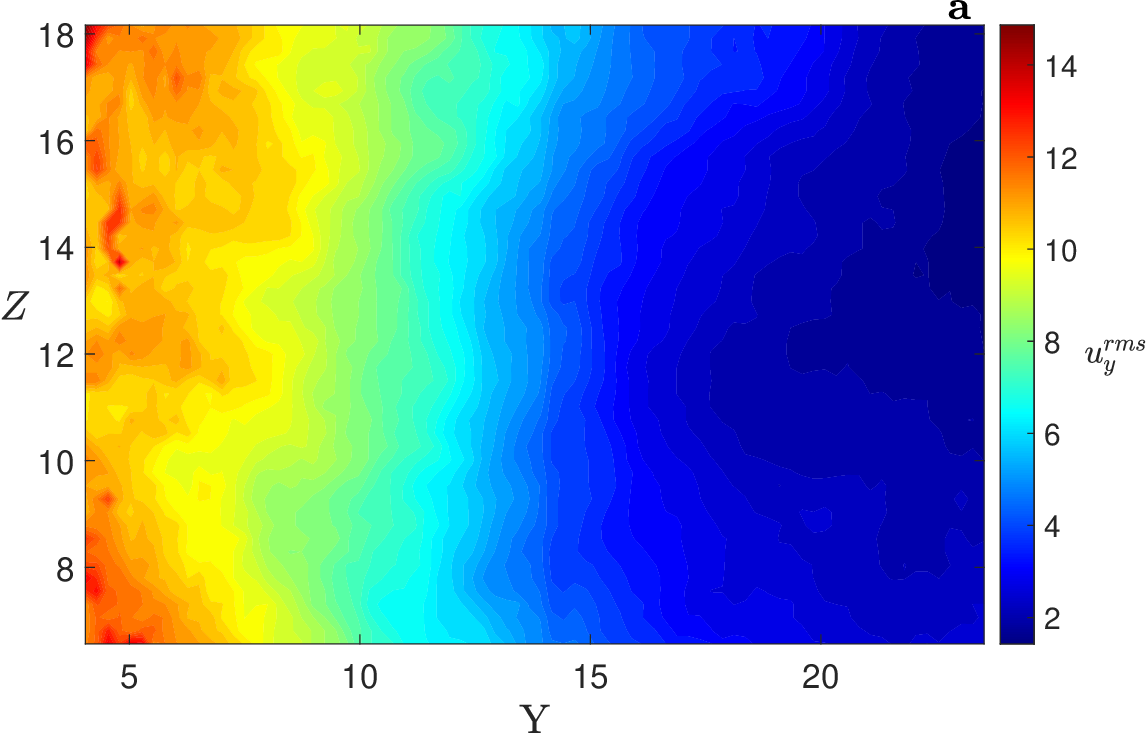}
\includegraphics[width=7.0cm]{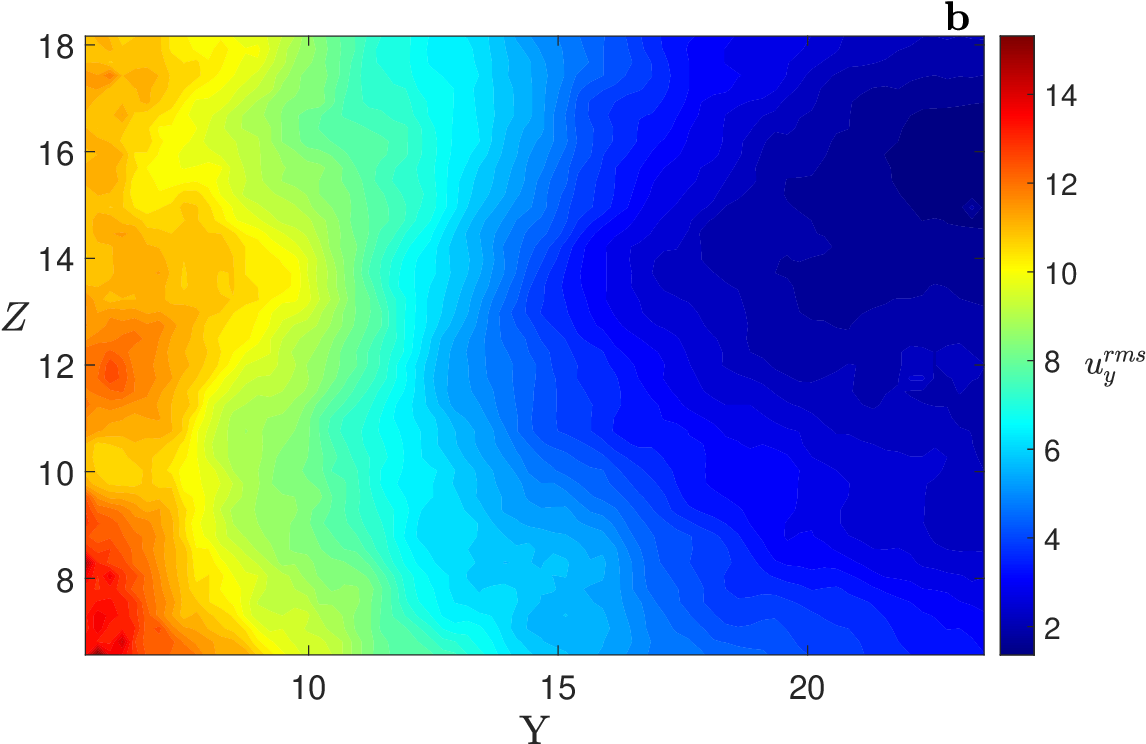}
\caption{\label{Fig8}
Counter  lines of the horizontal component of the turbulent velocity $u^{\rm (rms)}_y$
for {\bf (a)} isothermal turbulence (upper panel) and {\bf (b)} temperature stratified turbulence
(lower panel)  in the $YZ$ plane.
The velocity is measured in cm/s and coordinates  are in cm.
}
\end{figure}

\begin{figure}
\centering
\includegraphics[width=7.0cm]{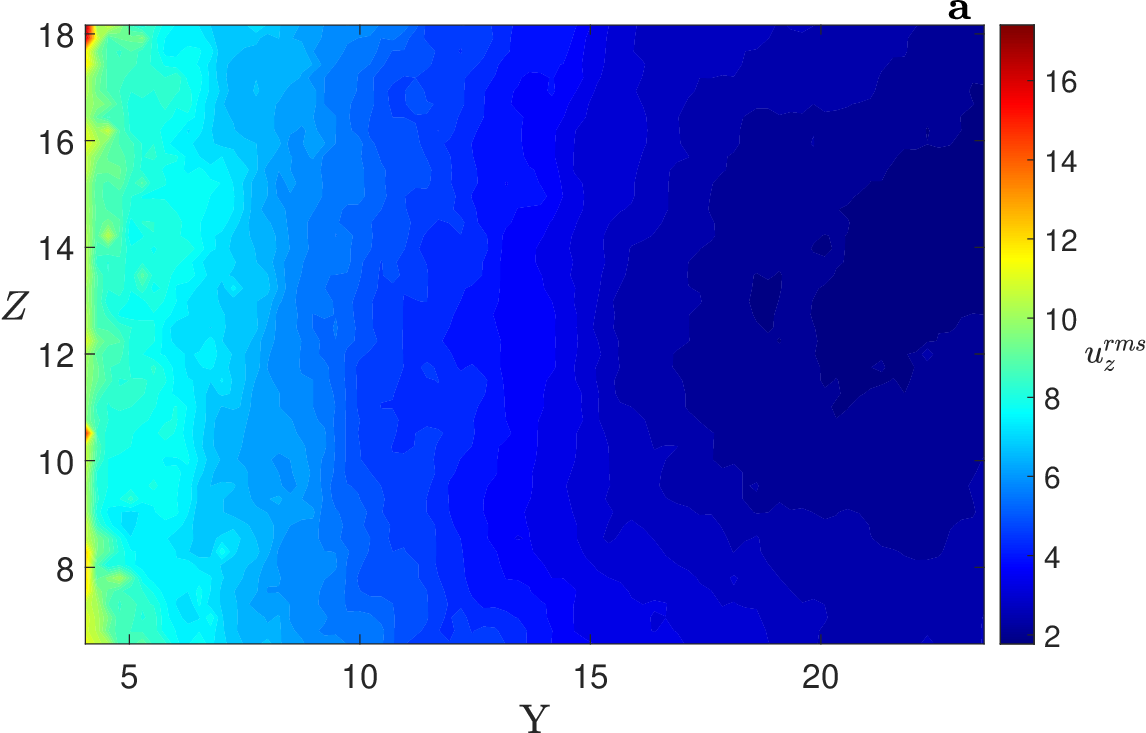}
\includegraphics[width=7.0cm]{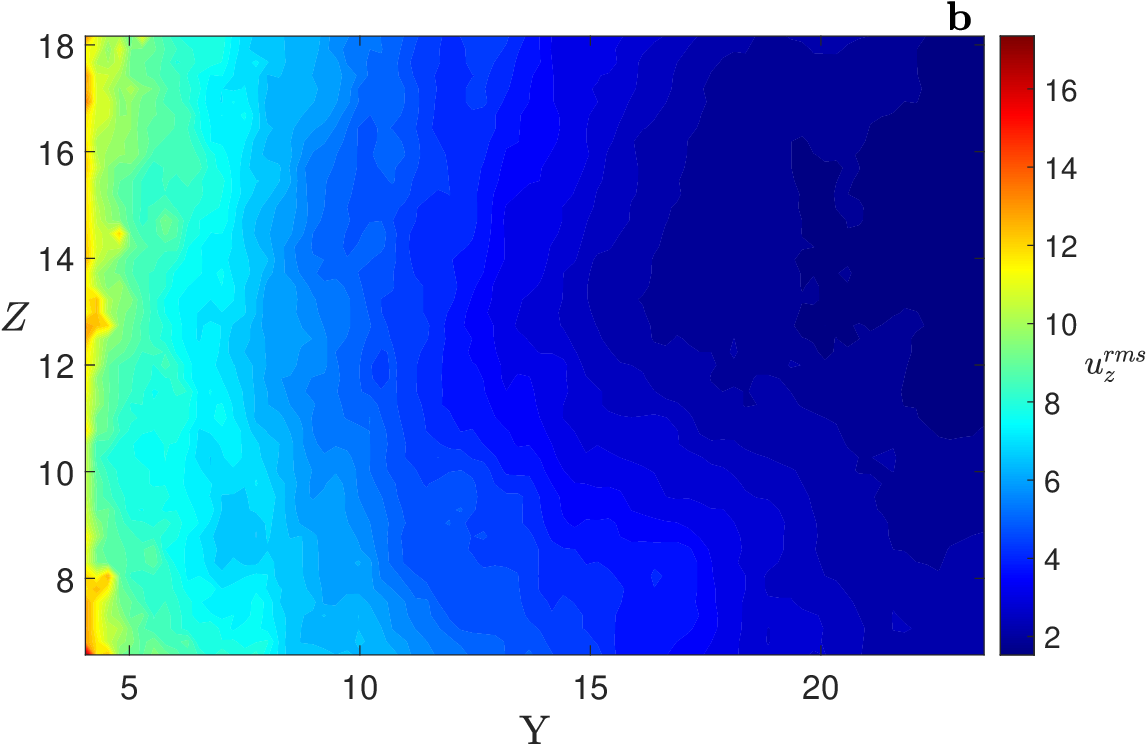}
\caption{\label{Fig9}
Counter lines for the vertical component of the turbulent velocity $u^{\rm (rms)}_z$
for {\bf (a)} isothermal turbulence (upper panel) and {\bf (b)} temperature stratified turbulence
(lower panel)  in the $YZ$ plane.
The velocity is measured in cm/s and coordinates  are in cm.
}
\end{figure}

\begin{figure}
\centering
\includegraphics[width=7.0cm]{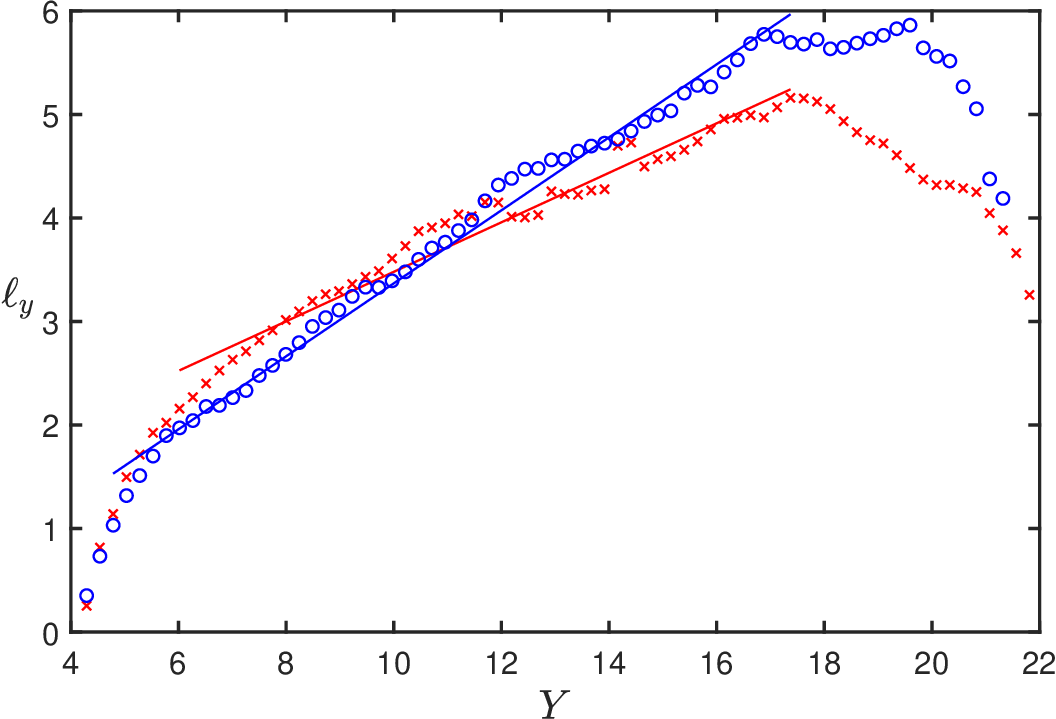}
\caption{\label{Fig10}
Horizontal integral turbulent scale $\ell_y$ versus $Y$ averaged
over $Z$ for isothermal turbulence (red); fitting:
$\ell_y= 0.24Y +0.96$ (solid) from $Y=$ 6 cm to 17.3 cm and temperature
stratified turbulence (blue); $\ell_y= 0.36Y - 0.33$
(solid) from $Y=$ 4.7 cm to 17.3 cm.
The integral scale and coordinates are measured in cm.
}
\end{figure}

\begin{figure}
\centering
\includegraphics[width=7.0cm]{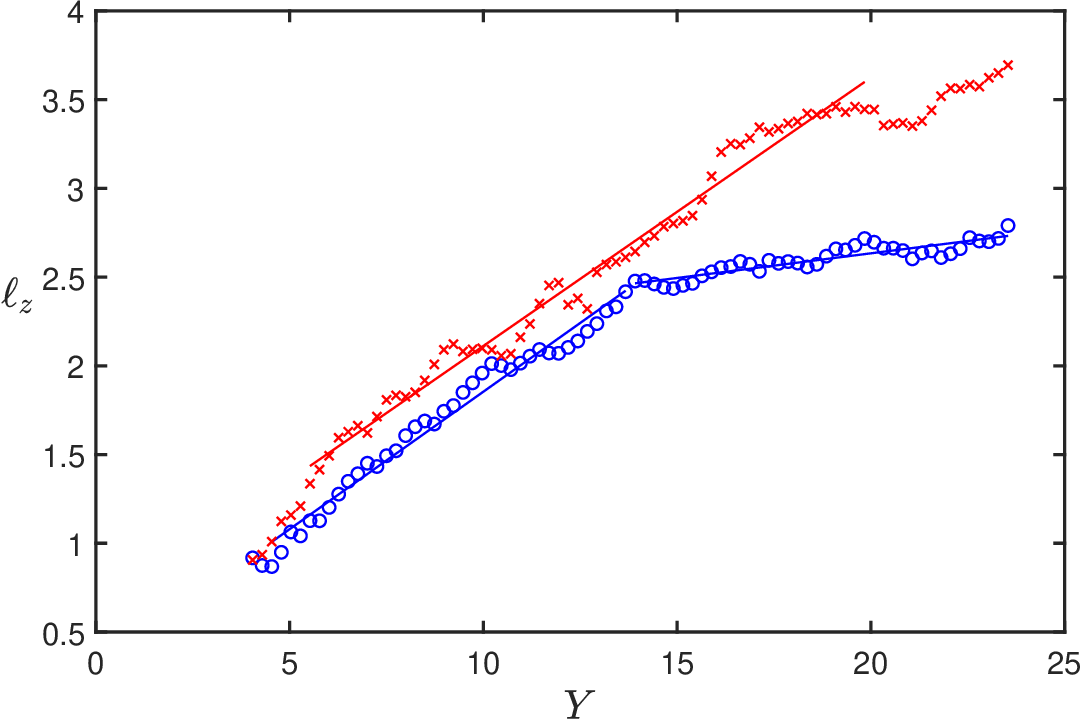}
\caption{\label{Fig11}
Vertical integral turbulent scale $\ell_z$ versus $Y$ averaged
over $Z$ for isothermal turbulence (red); fitting:
$\ell_z= 0.15Y + 0.6$ (solid) from $Y=$ 5.5 cm to 19.8 cm
and temperature stratified turbulence (blue); fitting:
$\ell_z= 0.155Y + 0.3$ (solid) from $Y=$ 4.5 cm to 13.5 cm and
$\ell_z= 0.027Y + 2.08$ from $Y=$ 13.5 cm to 23.7 cm.
The integral scale and coordinates are measured in cm.
}
\end{figure}

\begin{figure}
\centering
\includegraphics[width=7.0cm]{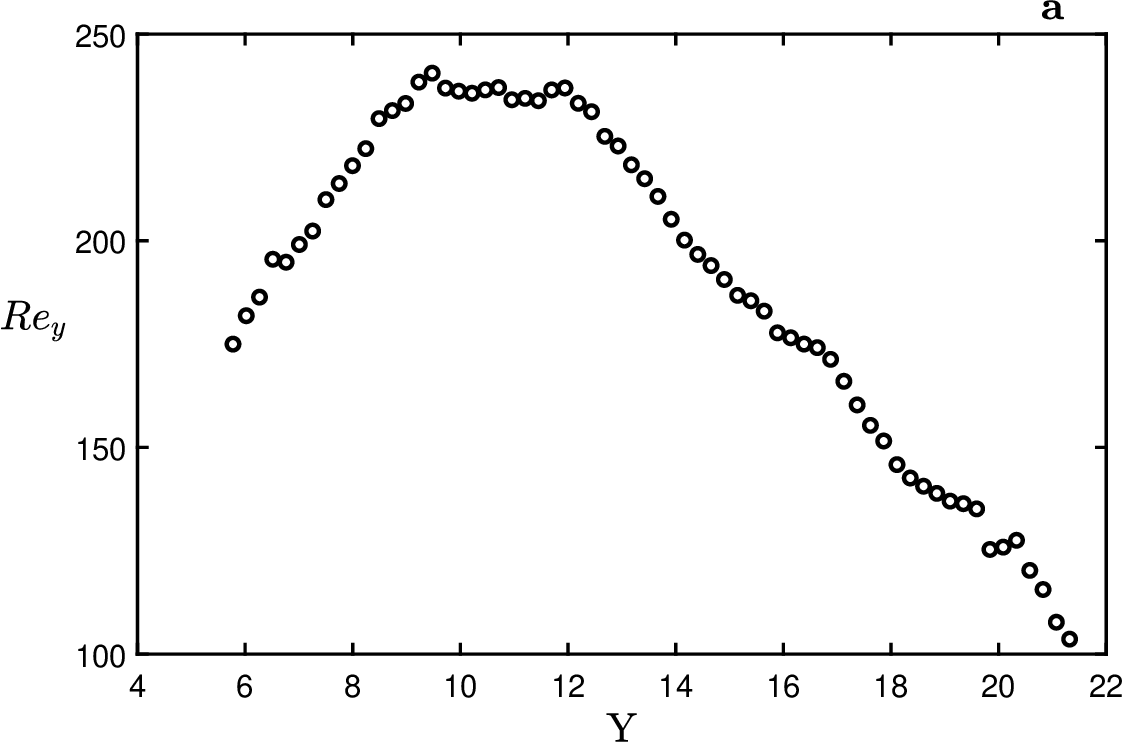}
\includegraphics[width=7.0cm]{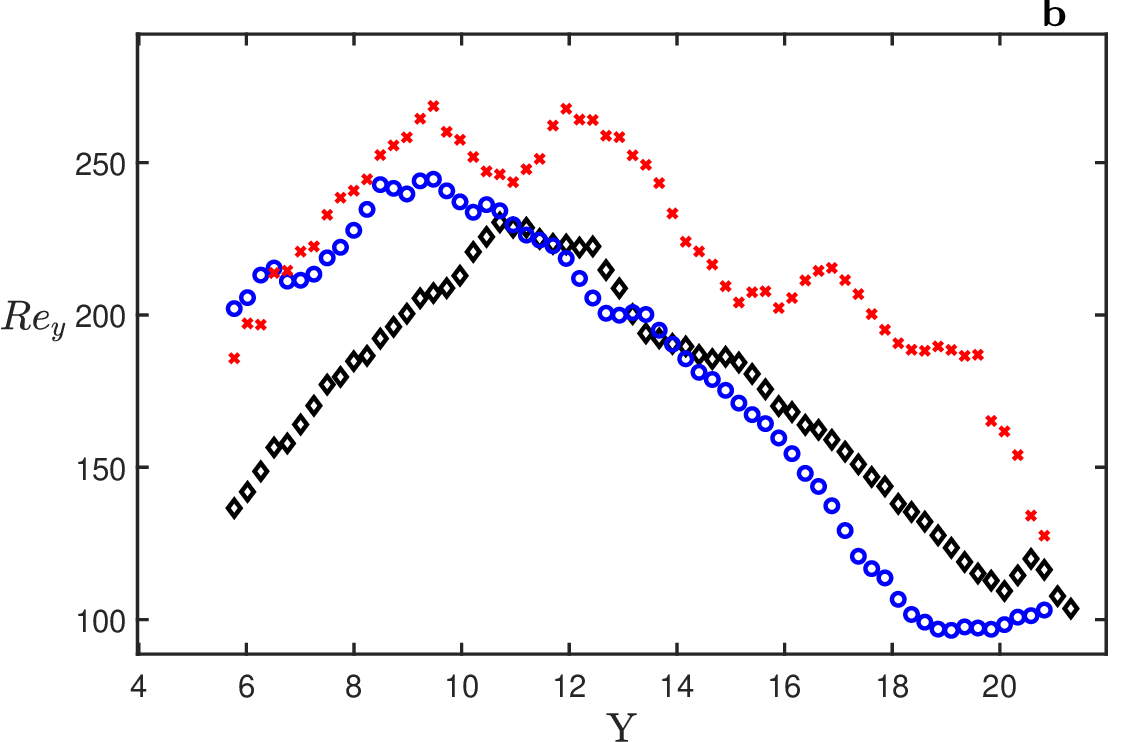}
\caption{\label{Fig12}
Reynolds number Re$_y = u^{\rm (rms)}_y \, \ell_y / \nu$ in the horizontal direction versus $Y$
{\bf (a)} averaged over $Z$ (upper panel) or {\bf (b)} averaged over different vertical regions (lower panel):
$Z=4.3-9.6$ cm (red, slanting crosses); $Z=9.6-15$ cm (blue, circles);
$Z=15-20$ cm (black, diamond) for temperature stratified turbulence.
}
\end{figure}

\begin{figure}
\centering
\includegraphics[width=7.0cm]{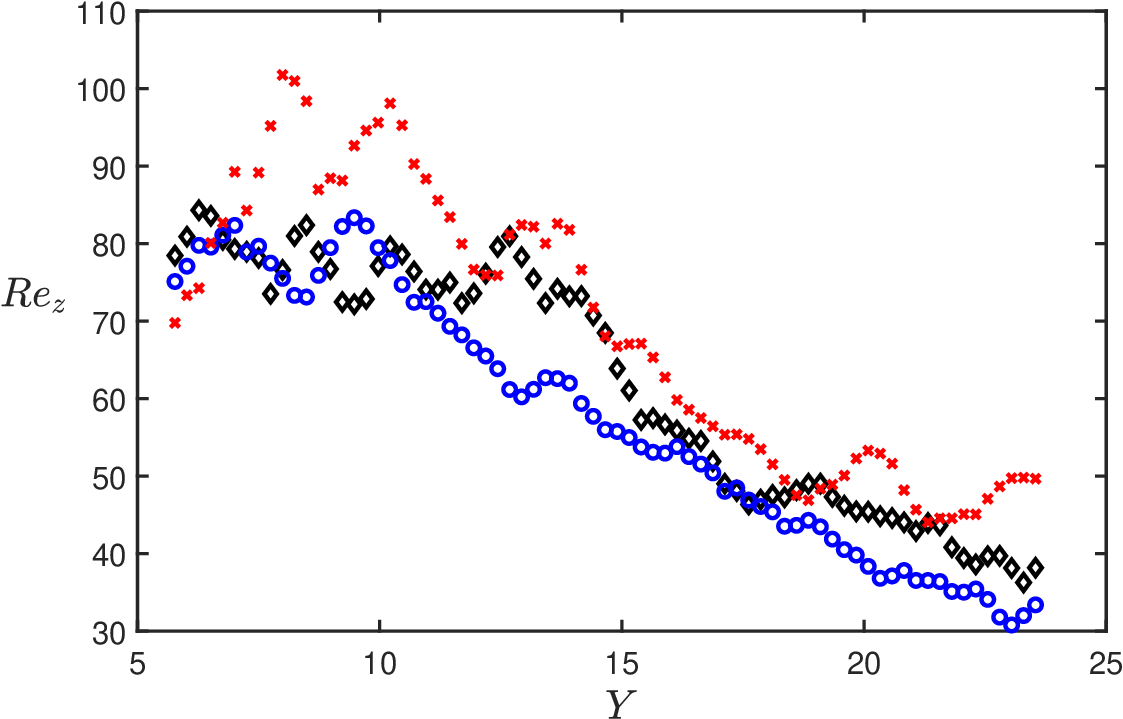}
\caption{\label{Fig13}
Reynolds number Re$_z = u^{\rm (rms)}_z \, \ell_z / \nu$ in the vertical direction versus $Y$
averaged over different vertical regions:  $Z=4.3-9.6$ cm (red, slanting crosses);
$Z=9.6-15$ cm (blue, circles) and $Z=15-20$ cm (black, diamond) for
temperature stratified turbulence.
}
\end{figure}

\begin{figure}
\centering
\includegraphics[width=7.0cm]{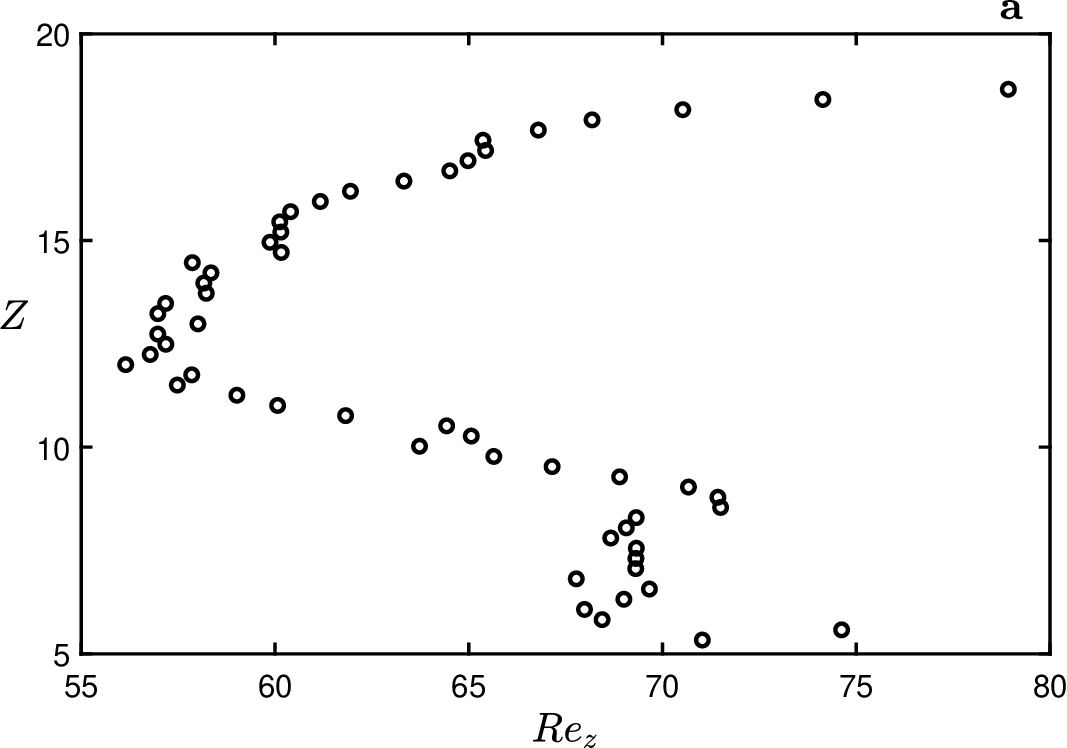}
\includegraphics[width=7.0cm]{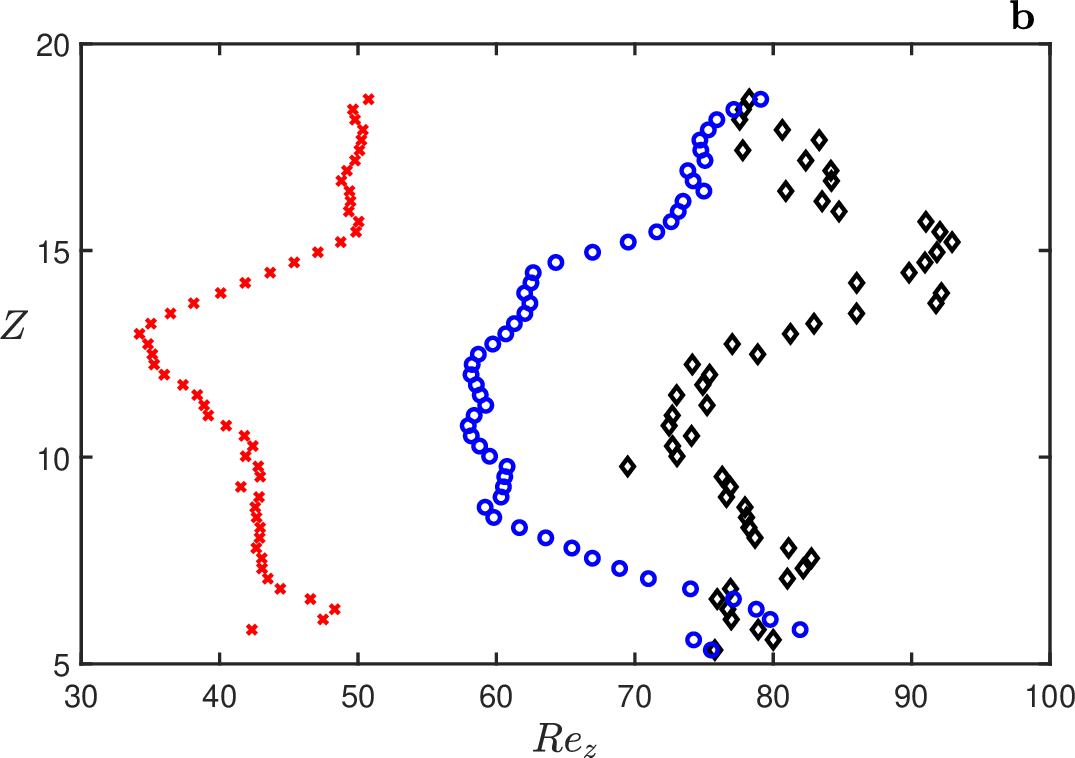}
\caption{\label{Fig14}
Reynolds number Re$_z = u^{\rm (rms)}_z \, \ell_z / \nu$ in the vertical direction versus $Z$
{\bf (a)} averaged over $Y$ (upper panel) or {\bf (b)} averaged over different horizontal regions (lower panel): $Y=4-10$ cm (black, diamond); $Y=10-17$ cm (blue, circles) and $Y=17-23$ cm (red, slanting crosses)
for temperature stratified turbulence.
}
\end{figure}

\begin{figure}
\centering
\includegraphics[width=8.0cm]{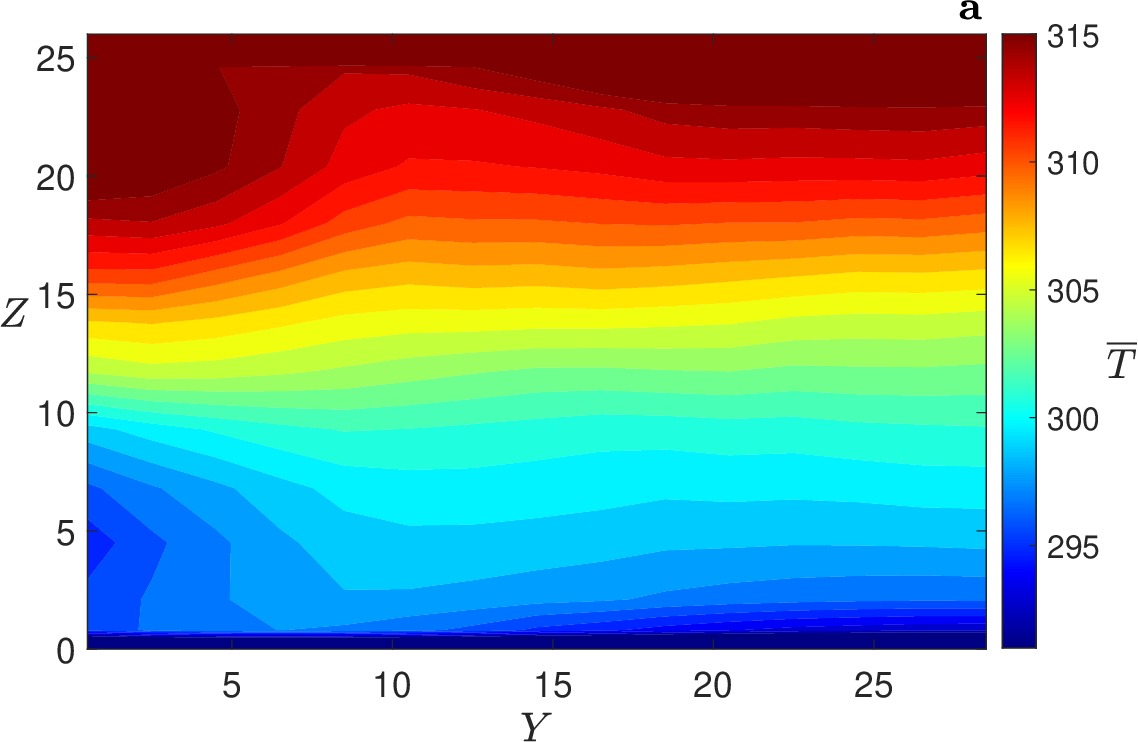}
\includegraphics[width=8.0cm]{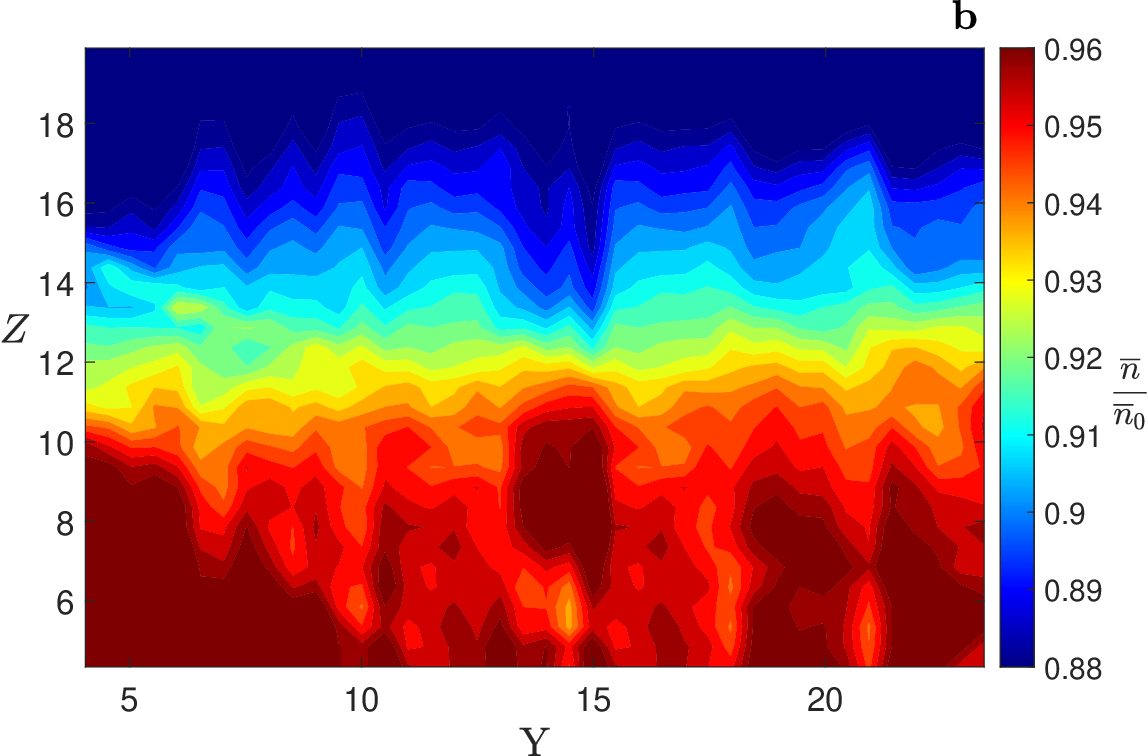}
\caption{\label{Fig15}
Distributions of {\bf (a)} the mean temperature $\meanT(Y,Z)$ (upper panel) and {\bf (b)} normalized mean particle number density $\meanN(Y,Z) / \meanN_0$ (lower panel)
for temperature stratified turbulence. Temperature is measured in K and coordinates are measured in cm.
}
\end{figure}

\begin{figure}
\centering
\includegraphics[width=8.0cm]{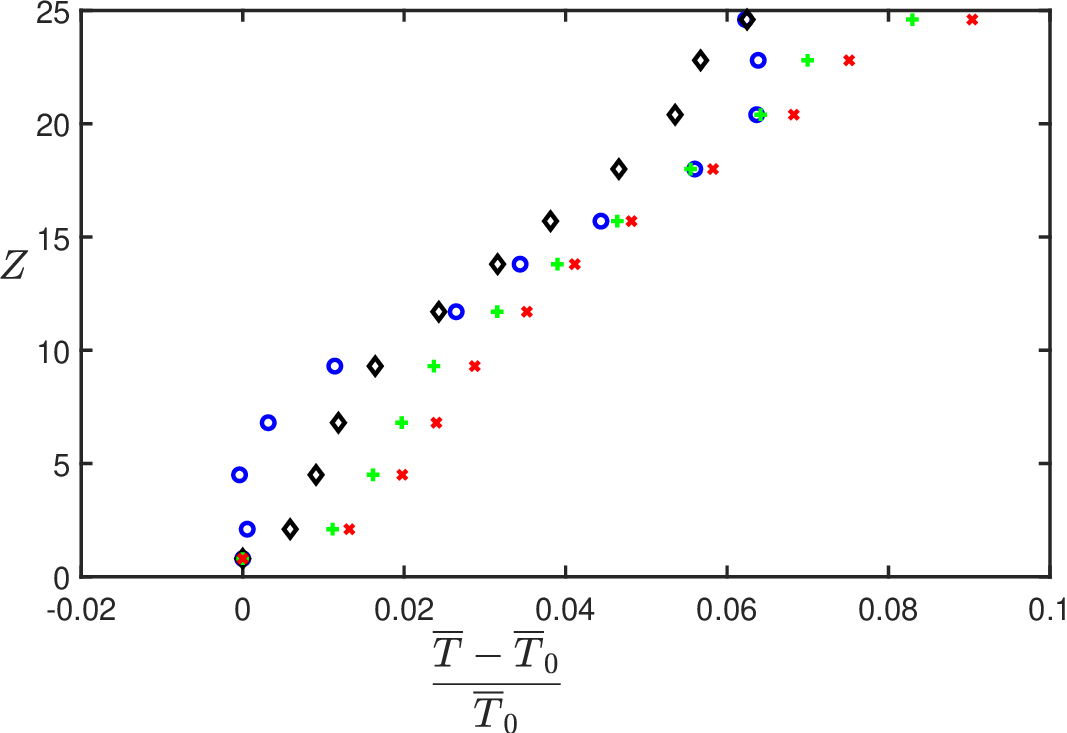}
\caption{\label{Fig16}
Vertical profiles of the relative normalized mean temperature $(\meanT - \meanT_0) / \meanT_0$
averaged over different horizontal regions:
$Y=0.5-6.5$ cm (blue, circles); $Y=8.5-14.5$ cm (black, diamond); $Y=16.5-22.5$ cm (green, crosses) and $Y=24.5-28.5$ cm (red, slanting crosses).
}
\end{figure}

\begin{figure}
\centering
\includegraphics[width=8.0cm]{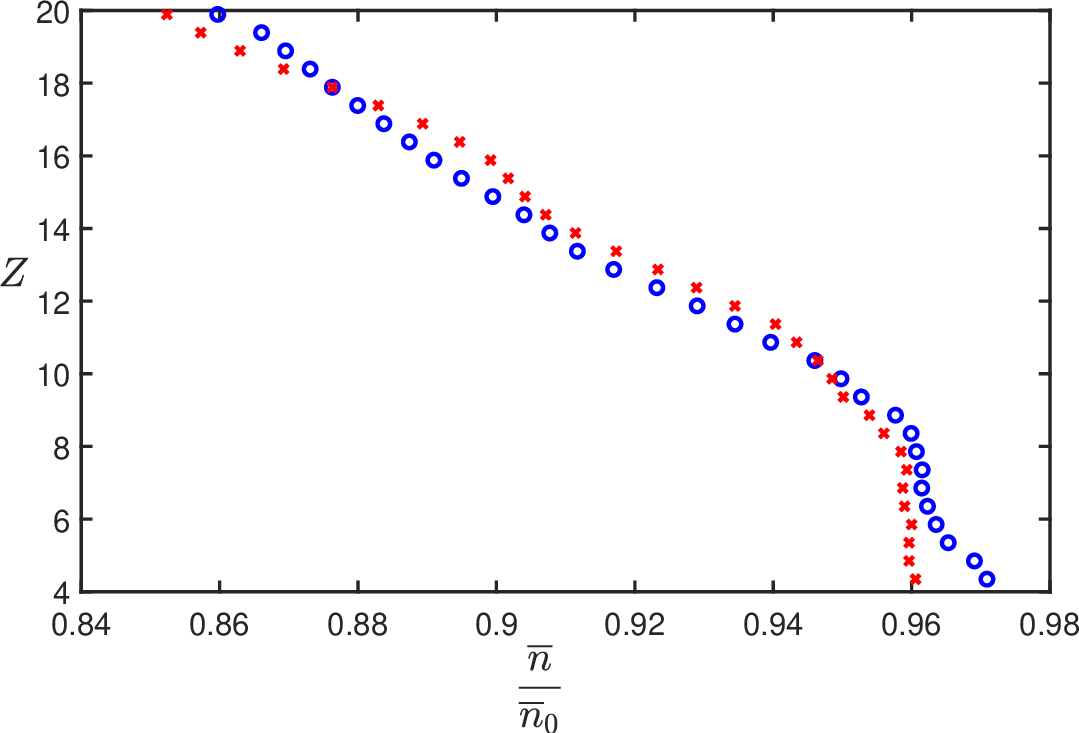}
\caption{\label{Fig17}
Vertical profiles of the normalized mean particle number density $\meanN / \meanN_0$
averaged over different horizontal regions:
$Y=4-15$ cm (blue, circles) and $Y=15-24$ cm (slanting crosses)
for temperature stratified turbulence.
}
\end{figure}

In Figs.~\ref{Fig2} and~\ref{Fig3} we show mean velocity patterns $\overline{U}$  in the core flow for isothermal turblulence and temperature stratified turbulence. All measurements of the velocity field in the horizontal direction $Y$ are performed starting 20 cm away from the left wall of the chamber where the grid is located. The amplitude of the grid oscillations is 6 cm, and velocity is measured beginning of the distance 4 cm away from the oscillating grid.
Figures~\ref{Fig2}and~\ref{Fig3} demonstrates that the nonuniform mean temperature field affects the mean velocity patterns.

In Figs.~\ref{Fig4} and~\ref{Fig5} we plot turbulent velocities $u^{\rm (rms)}_y$ and $u^{\rm (rms)}_z$ versus $Y$
averaged over different vertical regions (see for details, captions of Figs.~\ref{Fig4} and~\ref{Fig5})
for isothermal and temperature stratified turbulence.
In Fig.~\ref{Fig6} and~\ref{Fig7} we also show turbulent velocities $u^{\rm (rms)}_y$ and $u^{\rm (rms)}_z$ versus $Y$
averaged over $Z$  (without the separation to different vertical regions) for isothermal and temperature stratified turbulence.
The differences in the turbulent velocities for isothermal and temperature stratified turbulence
are not essential. This is also seen in Figs.~\ref{Fig8} and~\ref{Fig9} where we show the counter lines for the horizontal and vertical components of the turbulent velocity in the $YZ$ plane for isothermal and  temperature stratified turbulence.

\begin{figure}
\centering
\includegraphics[width=8.0cm]{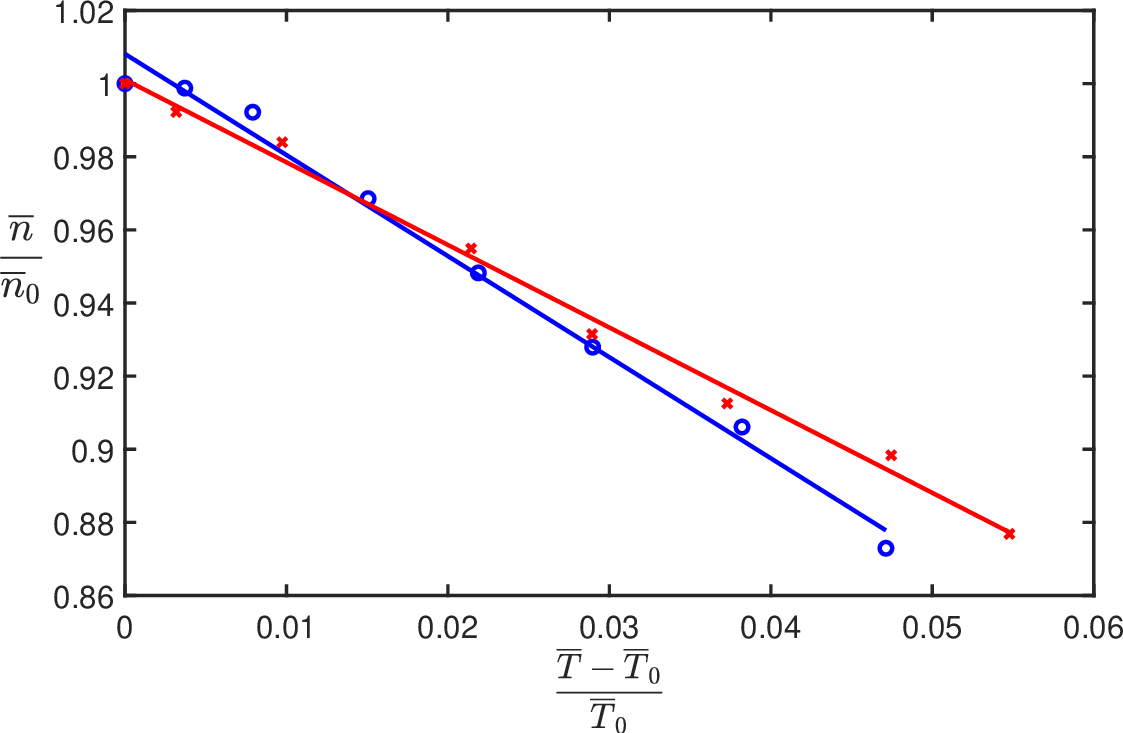}
\caption{\label{Fig18}
The normalized mean particle number density $\meanN / \meanN_0$ versus
the relative normalized mean temperature $(\meanT - \meanT_0) / \meanT_0$
averaged over different horizontal regions: $Y=4-15$ cm (blue, circles) and $Y=15-24$ cm (red, slanting crosses)
for temperature stratified turbulence.
}
\end{figure}

In Figs.~\ref{Fig10} and~\ref{Fig11} we plot the integral turbulent scales $\ell_y$  and $\ell_z$
in horizontal and vertical directions versus $Y$
averaged over $Z$ for isothermal and temperature stratified turbulence.
In both cases, in isothermal and temperature stratified turbulence,
the vertical component of the turbulent velocity $u^{\rm (rms)}_z$ decreases as $Y^{-1}$,
while the integral turbulent scales $\ell_y$  and $\ell_z$ increase linearly with the distance from the grid $Y$
in agreement with previous studies for isothermal turbulence in water experiments
\cite{turn68,turn73,tho75,hop76,kit97,san98,med01}.

Using the obtained  turbulent velocities $u^{\rm (rms)}_i$ and the integral turbulence scales $\ell_i$, we determine the Reynolds numbers in the horizontal $\left({\rm Re}_y = u^{\rm (rms)}_y \, \ell_y / \nu\right)$ and the vertical $\left({\rm Re}_z = u^{\rm (rms)}_z \, \ell_z / \nu\right)$ directions as functions of horizontal coordinate $Y$ (see Figs.~\ref{Fig12} and~\ref{Fig13}).
In a similar way, we determine the vertical profiles of Re$_z$ plotted in Fig.~\ref{Fig14}.
The Reynolds numbers  ${\rm Re}_y$ in the horizontal direction are essentially larger than the Reynolds numbers ${\rm Re}_z$ in the vertical direction, because the turbulent velocities and the integral turbulence scales are larger in the horizontal direction.

The Stokes number for particles used in our experiments
varies in the range from $5 \times 10^{-6}$ far from the grid to $2 \times 10^{-4}$ near the grid.
In particular, the Stokes time for particles having the diameter 0.7 $\mu$m is $1.5 \times 10^{-6}$ s,
and the Kolmogorov time varies from $7 \times 10^{-3}$ s in turbulence near the grid
to $0.3$ s in turbulence far from the grid.

The initial distribution of particles injected into the chamber is nearly homogeneous.
Turbulent thermal diffusion results in formation of strongly inhomogeneous particle distributions.
Sedimentation of particles is very weak in our experiments, since the terminal fall velocity for
particles having the mean diameter $0.7 \mu$m is about $10^{-2}$ cm/s.
Characteristic turbulent velocity (that is about $12$ cm/s near the grid and is about $2$ cm/s far from the grid)
and the effective pumping velocity caused by turbulent thermal diffusion (that is about $1$ cm/s near
the grid and is about $0.1$ cm/s far from the grid)
are much larger than the terminal fall velocity.

Measurements of temperature and particle number density allow us to determine
spatial distribution of the mean temperature $\meanT$
and the normalized mean particle number density $\meanN / \meanN_0$ (see Fig.~\ref{Fig15}).
Inspection of Fig.~\ref{Fig15} shows that particles are accumulated in the vicinity
of the minimum of the mean temperature (i.e., in the vicinity of the bottom wall of the chamber)
due to phenomenon of turbulent thermal diffusion.

In Fig.~\ref{Fig16} we show vertical profiles of the relative normalized mean temperature
$(\meanT - \meanT_0) / \meanT_0$ averaged over different horizontal regions
(see the caption of Fig.~\ref{Fig16}), where $\meanT_0$ is the reference mean temperature.
In Fig.~\ref{Fig17} we plot the normalized mean particle number density
$\meanN / \meanN_0$ averaged over different horizontal regions (see the caption of Fig.~\ref{Fig17}),
where  $\meanN_0=\meanN(\meanT = \meanT_0)$.
The normalized mean temperature
increases linearly with the height $Z$ in the flow core (see Fig.~\ref{Fig16}),
while the normalized mean particle number density
decreases linearly with the height $Z$ (see Fig.~\ref{Fig17}).

The vertical profiles of the mean temperature and the mean particle number density
allow us to determine the normalized mean particle number density
$\meanN / \meanN_0$ versus the relative normalized mean temperature $(\meanT - \meanT_0) / \meanT_0$
(see Fig.~\ref{Fig18}).
The slope of this dependence allows us to find the effective turbulent thermal diffusion coefficient
$\alpha$ for particles in inhomogeneous and anisotropic stably stratified turbulence.
Indeed, the steady-state solution~(\ref{D12})  of the equation for the mean particle number density
yields $\delta\meanN / \meanN_0 = - \alpha \, \delta\meanT  / \meanT_0$,
where we take into account that $D_{\rm T} \gg D$.
This equation can be rewritten as $\meanN / \meanN_0 = 1 - \alpha \, (\meanT - \meanT_0) / \meanT_0$.
Using this equation, we find from Fig.~\ref{Fig13}, that $\alpha=2.765$ for particles accumulated in the regions $Y=4-15$ cm,
and $\alpha=2.26$ for particles accumulated in the regions $Y=16-24$ cm.
For the second region that is far from the grid, the level of turbulence is lower than that for the first region.
This explains the difference in the effective turbulent thermal diffusion coefficient
$\alpha$ determined for these regions.
This experimental study has clearly detected phenomenon of turbulent thermal diffusion
in an inhomogeneous turbulence.

\section{Conclusions}
\label{sect5}

Turbulent thermal diffusion of particles in inhomogeneous stably stratified turbulence
produced by one oscillating grid in the air flow has been studied.
Measurements of the velocity fields using a Particle Image Velocimetry (PIV)
allow us to determine the mean and the r.m.s. velocities,
two-point correlation functions of the velocity field and an integral scale of turbulence.
Spatial distributions of the temperature field have been determined using a temperature probe equipped with 12 E – thermocouples.
We also determine the spatial distributions of particles by means of the PIV system using the effect of the Mie light scattering
by particles in the flow.
The experiments have shown that particles are accumulated at the vicinity of the minimum of the mean temperature
due to the phenomenon of turbulent thermal diffusion.
The obtained spatial distributions of particles and temperature fields
allow us to determine the effective turbulent thermal diffusion coefficient
of particles in inhomogeneous temperature stratified turbulence.
This coefficient varies from $2.765$ for particles accumulated in the vicinity of the cold wall
of the chamber in the regions more close to the grid to $2.26$ for particles accumulated in the regions far from the grid
where turbulence is less intensive. These values are in agreement with theoretical predictions \cite{AEKR17}
for the micron-size particles.

\begin{acknowledgements}
This research was supported in part by the Israel Ministry of Science and Technology (grant No. 3-16516).
\end{acknowledgements}

\bigskip
{\bf DATA AVAILABILITY}
\medskip

The data that support the findings of this study are available from the corresponding author
upon reasonable request.

\end{document}